\documentclass[12pt, draftclsnofoot, onecolumn]{IEEEtran}
\ifCLASSINFOpdf
 \usepackage[pdftex]{graphicx}
\else
  \usepackage[dvips]{graphicx}
\fi

\usepackage{algorithm} %format of the algorithm
\usepackage{multirow} %multirow for format of table
\usepackage{xcolor}
\usepackage{algpseudocode}

\usepackage{graphicx}
\usepackage[cmex10]{amsmath}
\usepackage{amsfonts}
\usepackage{amssymb}
\usepackage{amsthm}
\usepackage{morefloats}
\usepackage{cite}
\usepackage{bm}
\usepackage{stfloats}

\usepackage[caption=false,font=normalsize,labelfont=sf,textfont=sf]{subfig}
\newtheorem{theorem}{Theorem}
\IEEEoverridecommandlockouts

\begin{document}

% can use linebreaks \\ within to get better formatting as desired
\title{Clustered Cell-Free Networking: A Graph Partitioning Approach}
%\title{RC-NetDecomp: Rate-Constrained Network Decomposition for Clustered Cell-Free Networking}
\author{Junyuan~Wang,~\IEEEmembership{Member,~IEEE},~Lin~Dai,~\IEEEmembership{Senior~Member,~IEEE},\\~Lu~Yang,~\IEEEmembership{Member,~IEEE},~and~Bo~Bai~\IEEEmembership{Senior~Member,~IEEE}
\thanks{This paper was presented in part in IEEE International Conference on Communications (ICC), Seoul, South Korea, May 2022 \cite{Conference}.}
%\thanks{Manuscript received March 6, 2017; revised July 31, 2017; November 28, 2017; accepted December 30, 2017. This paper will be presented in part at the IEEE International Conference on Communications (ICC), Kansas City, May 2018. The associate editor coordinating the review of this paper and approving it for publication was E. Jorswieck.}
\thanks{J. Wang is with the College of Electronic and Information Engineering and the Institute of Advanced Study, Tongji University, Shanghai, China (email: junyuanwang@tongji.edu.cn).}
\thanks{L. Dai is with the Department of Electrical Engineering, City University of Hong Kong, Hong Kong SAR, China (email: lindai@cityu.edu.hk).}
\thanks{L. Yang and B. Bai are with the Theory Lab, CRI, 2012 Labs, Huawei Technologies Co. Ltd., Hong Kong SAR, China (e-mail: yanglu87@huawei.com, baibo8@huawei.com).}}

%\author{Junyuan~Wang$^{*}$, Yuan~Kai$^{\dagger}$ and Huiling~Zhu$^{\dag}$ \\
%	$^{*}$Department of Computer Science, Edge Hill University, United Kingdom\\
%	$^{\dagger}$School of Engineering and Digital Arts, University of Kent, United Kingdom \\
%	Email: junyuan.wang@edgehill.ac.uk, yk69@kent.ac.uk, h.zhu@kent.ac.uk}

\maketitle
\vspace{-10mm}
\begin{abstract}
By moving to millimeter wave (mmWave) frequencies, base stations (BSs) will be densely deployed to provide seamless coverage in sixth generation (6G) mobile communication systems, which, unfortunately, leads to severe cell-edge problem. In addition, with massive multiple-input-multiple-output (MIMO) antenna arrays employed at BSs, the beamspace channel is sparse for each user, and thus there is no need to serve all the users in a cell by all the beams therein jointly. Therefore, it is of paramount importance to develop a flexible clustered cell-free networking scheme that can decompose the whole network into a number of weakly interfered small subnetworks operating independently and in parallel. Given a per-user rate constraint for service quality guarantee, this paper aims to maximize the number of decomposed subnetworks so as to reduce the signaling overhead and system complexity as much as   possible. By formulating it as a bipartite graph partitioning problem, a rate-constrained network decomposition (RC-NetDecomp) algorithm is proposed, which can smoothly tune the network structure from the current cellular network with simple beam allocation to a fully cooperative network by increasing the required per-user rate. Simulation results demonstrate that the proposed RC-NetDecomp algorithm outperforms existing baselines in terms of average per-user rate, fairness among users and energy efficiency.
\end{abstract}

\vspace{-5mm}

\vspace{-4mm}
\section{Introduction}
Mobile communication system has been evolving from its first generation (1G) to the current fifth generation (5G) under the cellular network structure over decades, where each base-station (BS) covers a given area and serves the users within its coverage independently. In order to meet the ever-increasing high data rate requirements, it has been proposed to explore the millimeter wave (mmWave) bands and even the terahertz bands in sixth generation (6G) mobile communication systems \cite{6G}. As mmWave/terahertz signals suffer from high path-loss and blockage effect, the coverage of a BS could reduce to a few hundreds or even tens of meters, leading to dense deployment of BSs for the sake of seamless coverage. It could be then expected that a large number of users would be located in the cell-edge areas and thus suffer from severe interference from neighboring BSs under the cellular network structure. Owing to the escalated cell-edge problem and the motivation to serve a user by multiple BSs to avoid signal blockage, we have to rethink whether the current cellular network structure is still suitable for 6G mobile communications.

\subsection{Cell-Free Massive MIMO}
An alternative network topology is the large-scale distributed antenna system (DAS) \cite{Zhuang, JZhang1, DAS_Book, Dai_JSAC, Zhiyang, Junyuan_DAS}, where a large number of geographically distributed access points (APs)\footnote{Note that the AP here can be regarded as a mini BS. In this paper, AP and BS will be used interchangeably.} are connected to a central processing unit (CPU) via optical fibers and jointly serve all the users in the system. The DAS has been researched under the cloud radio access network (C-RAN) architecture \cite{CRAN}, and was recently redefined as cell-free massive multiple-input-multiple-output (MIMO) in \cite{Ngo_SmallCell} by introducing a simple conjugate beamforming scheme. With conjugate beamforming, each AP estimates the channel state information (CSI) from all the users to it, and then applies conjugate beamforming to serve the users independently. Though without frequent CSI exchange among APs, the number of channels that need to be measured increases linearly with both the number of APs and the number of users, resulting in huge CSI signaling overhead and high signal processing complexity in cell-free massive MIMO systems.     

In fact, a user's signal power is mainly contributed by a number of surrounding APs, as the path-loss increases exponentially with the access distance from a user to an AP, indicating that coordinating all the APs to jointly serve all the users is unnecessary. A user-centric \emph{virtual-cell} concept was first proposed in \cite{Dai_Globecom, Dai_CDMA} in 2002. Specifically, each user selects a few APs with the highest channel gains to form its own virtual cell and a user is served by the APs within its virtual cell only. Such a user-centric approach is gaining increasing momentum due to the AP densification in 5G systems and beyond \cite{Dai_VC,Junyuan_VC,Ngo_VC,Lozano1,Lozano2,Buzzi1,Buzzi2,Hanzo_VC1,Hanzo_VC2,Jemin,Larsson,JZhang}. A user-centric virtual cell is mainly formed in four approaches: 1) each user selects a given number of APs with the best channel qualities or smallest distances \cite{Dai_VC,Junyuan_VC,Ngo_VC,Lozano1,Lozano2}; 2) each AP serves a given number of users with the best channels \cite{Buzzi1,Buzzi2}; 3) each AP serves the users within its coverage of some radius \cite{Hanzo_VC1,Hanzo_VC2,Jemin}; and 4) each user selects the APs with the best channels that contribute a given percentage to the overall channel gain \cite{Larsson,JZhang}.

Though simple, the user-centric virtual cells formed by the above methods are usually overlapped with each other, i.e., an AP could belong to multiple virtual cells, which complicates the downlink transmissions as the virtual cells are coupled with each other by the per-AP power constraint. Moreover, although a user is always served by its surrounding APs in the user-centric approach with cell-edge problem avoided, it still suffers from interference from other virtual cells, which limits the system spectral efficiency. A number of research work was then devoted to the interference management problems in user-centric virtual-cell based networks. For instance, power control algorithms were studied in \cite{Ngo_VC,Buzzi2}. A local minimum mean square error (MMSE) receiver was proposed in \cite{Lozano1}, and a parallel interference cancellation scheme was further designed in \cite{Lozano2} to reduce the complexity of uplink signal reception. For downlink transmission, a local partial zero-forcing precoding scheme was proposed in \cite{Larsson}. By introducing orthogonality among virtual cells, \cite{Hanzo_VC1} and \cite{JZhang} studied the resource block allocation problem and the pilot assignment problem, respectively, based on the graph coloring theory.

\subsection{Clustered Cell-Free Networking}
Since the aforementioned approaches optimize the data transmission of virtual cells jointly to manage interference \cite{Ngo_VC, Buzzi2,Lozano1,Lozano2,Larsson}, huge CSI exchange and high computational complexity are inevitable. A more practical network structure to facilitate efficient interference cancellation with tolerable complexity is to decompose the whole network into a number of small subnetworks such that the users strongly interfered with each other are within the same subnetwork and little interference exists among different subnetworks \cite{Lin}. In this way, efficient interference cancellation techniques can be applied within each subnetwork independently, and the inter-subnetwork interference could be treated as noise. Such a flexible networking scheme is referred to as \textit{clustered cell-free networking} in this paper. 

There are various ways to cluster users and APs into subnetworks. The previously proposed coordinated multi-point (CoMP) transmission \cite{MKarakayali,JZhang4,HDahrouj,KHuang} in cellular networks, for instance, can be regarded as an AP-centric clustering scheme, where the APs are first partitioned into multiple clusters and then each user joins the cluster its associated AP belongs to. Various dynamic AP clustering methods were proposed by adopting heuristic methods \cite{Gesbert,Saad,NLee,Larsson2,Goldsmith}. A representative scheme recently proposed in \cite{Goldsmith} adopted a hierarchical clustering algorithm to group APs based on the minimax linkage criterion \cite{minmax}. However, with AP-centric clustering algorithms, since the clusters are formed from the AP side only, there could be unlucky users located close to the cluster edge and hence suffer from strong interference from neighboring subnetworks \cite{KHuang}. 

In order to avoid the cluster-edge problem, user-centric clustering algorithms were proposed \cite{Junyuan_VC,Xiao,LChen,YLin}. Specifically, each user forms its own virtual cell by selecting a few nearby APs, and then virtual cells are merged to form subnetworks according to various criteria. For instance, \cite{Junyuan_VC} proposed to merge the virtual cells that share at least one AP. As a user is always located at the center of its own virtual cell, there are no cluster-edge users. However, the resulted subnetworks are usually imbalanced with a giant subnetwork as demonstrated in \cite{Lin}, implying that the joint processing complexity and signaling overhead could be still high after clustering. 

Note that in the above existing work on clustering users and APs, single-antenna APs are usually assumed. As massive MIMO antenna arrays would be widely deployed in 5G systems and beyond \cite{Femenias,Poor}, it is of paramount importance to study the clustered cell-free networking problem in a general cell-free wireless network with multiple APs, each equipped with a massive MIMO antenna array. In a massive MIMO system with preformed narrow and directional beams, the beamspace channel from a user to the beams is sparse and a user's beamspace channel is mainly determined by a few beams. Since only users with highly correlated beamspace channels are strongly interfered with each other \cite{Adhikary}, in order to reduce signaling overhead and system complexity, the beams and users can be grouped into a number of subnetworks with little inter-subnetwork interference. A beam-user clustering algorithm was proposed in \cite{SheuJS} in a single-cell setting, which first selects a fixed number of beams for each user with the highest beam gains and then groups strongly interfered users along with their associated beams together for intra-subnetwork interference cancellation. This is essentially a user-centric clustering algorithm similar to that proposed in \cite{Junyuan_VC}. 

In contrast to the aforementioned clustering algorithms which are either AP-centric or user-centric, it was recently proposed in \cite{Lin} that the network decomposition problem can be formulated as a bipartite graph partitioning problem by modeling the wireless network as a weighted undirected bipartite graph with the edge weight between an AP and a user defined as the corresponding channel gain. The proposed algorithm can significantly reduce inter-subnetwork interference compared to the AP-centric ones, and produce balanced subnetworks in the meantime. There are, nevertheless, a few key issues to be further addressed. First of all, the algorithm proposed in \cite{Lin} was under the constraint that the sum of the sum-interference-to-sum-signal ratios of all the subnetworks is smaller than some predefined threshold, while no rate performance guarantee of users was provided. In addition, it was required in \cite{Lin} that each subnetwork contains at least one user and one AP. As a result, all the APs are always involved in data transmission, which may lead to significant waste of energy in future dense mobile communication systems. Intuitively, the APs with no users nearby should be switched into sleep mode for power saving. Furthermore, single-antenna APs were assumed in \cite{Lin}, while massive MIMO antenna arrays would be widely deployed at the APs in 5G/6G mobile communication systems.  

\subsection{Our Contributions}
In this paper, the clustered cell-free networking problem is studied with new practical considerations including: 1) massive MIMO APs featured with dynamic on-off switch of beams,\footnote{An AP is switched into sleep mode if all the beams belonging to it are turned off.} and 2) a per-user rate constraint imposed for service quality guarantee. By considering a general multi-AP multi-beam system, where $L$ massive MIMO APs are deployed with a large number of preformed fixed beams available at each AP, the cell-free massive MIMO system with single-antenna APs and the single-cell massive MIMO system with fixed beams are included as two special cases.

For clustered cell-free networking, the whole network is decomposed into a number of small subnetworks operating independently and in parallel, with joint processing conducted only inside each subnetwork. Therefore, with more subnetworks, the subnetwork size becomes smaller on average, leading to lower signaling overhead and complexity of joint processing. Yet the intersubnetwork interference may increase, causing rate degradation. It is clear that the number of subnetworks determines a crucial tradeoff between rate performance and complexity/signaling overhead. In this paper, given a per-user rate constraint for service quality guarantee, we aim to decompose a large-scale network into as many subnetworks as possible, so as to reduce the joint processing complexity and signaling overhead to the most extent.\footnote{Note that an alternative problem formulation is to optimize the rate performance for a given constraint of complexity, such as the number of subnetworks or the maximum subnetwork size. Optimizing the rate performance could provide the best services by operating the system at the maximum allowed complexity, while maximizing the number of subnetworks for a given rate constraint could keep the system complexity to a minimum as long as the rate requirement is met. In this paper, we focus on the latter as it aims to minimize the joint processing complexity and signaling overhead as much as possible. It would be of great importance to further study the rate optimization problem for a given constraint of complexity in the future.} 

Specifically, to incorporate massive MIMO APs, a weighted undirected bipartite graph is constructed based on the beams at APs and users. To further enable on-off switch of beams, a novel edge weight model is proposed. Instead of modeling the edge weights between users and beams as the corresponding channel gains as \cite{Lin} did, in this paper, the edge weights incident to a user are modeled as the corresponding beamspace channel gains normalized by the highest channel gain between this user and all beams. As a result, each user has similar connectivity on the graph and is always incident to an edge with weight 1, indicating that all the users would be treated fairly. By contrast, the edge weights incident to different beams could be of significant difference, as the edge weights of a beam would be very small when it is badly aligned with users or its associated AP is far from users. As a result, isolating the beams with small edge weights, i.e., switching them off, would be prioritized. 
	  
Based on the proposed bipartite graph with the novel edge weight model, the clustered cell-free networking problem of maximizing the number of decomposed subnetworks with a per-user rate constraint is transformed to a graph partitioning problem with a graph cut based constraint. A novel rate-constrained network decomposition (RC-NetDecomp) algorithm is then proposed to solve the problem. Simulation results show that the proposed RC-NetDecomp algorithm achieves superior performance over the AP-centric baseline in terms of average per-user rate, user fairness and energy efficiency. Compared to the user-centric baseline, our RC-NetDecomp algorithm enables a flexible fine-grained tuning of the number of decomposed subnetworks and the subnetwork size. Moreover, for a given number of decomposed subnetworks, the proposed RC-NetDecomp algorithm produces more balanced subnetworks with smaller maximum subnetwork size than the user-centric approach, which corroborates its ability to efficiently reduce joint processing complexity and signaling overhead.

The remainder of this paper is organized as follows. Section II introduces the clustered cell-free network model and the rate-constrained network decomposition problem. A graph partitioning based clustered cell-free networking algorithm is proposed in Section III, followed by simulation results in Section IV. Finally, concluding remarks are summarized in Section V. 

Throughout this paper, $G=(V, E)$ denotes a graph with vertex set $V$ and edge set $E$. $\mathbb{E}[\cdot]$ denotes the expectation operator. The superscript $T$ denotes the transpose. $\mathbf{0}_{M\times N} $ denotes an $M \times N$ matrix with all entries zero. $x_{i,j}$ denotes the $i$th row and $j$th column entry of matrix $\mathbf{X}$. $|X|$ denotes the cardinality of set $X$. $X\cap Y$ and $X\cup Y$ denote the intersection and union of set $X$ and set $Y$, respectively.  $X \setminus Y$ denotes the relative complement of set $Y$ in set $X$. $\emptyset$ denotes the empty set.

\section{Clustered Cell-Free Network Model and Problem Formulation}

\subsection{Network Model and Bipartite Graph Representation}
Consider a general cell-free wireless network with $K$ single-antenna users and $L$ APs. It is assumed that each AP $l$ is equipped with a massive MIMO antenna array with $N_{l}$ fixed analog beams preformed. The total number of beams $N$ in the network is then given by $N=\sum_{l=1}^{L}N_{l}$. Let us denote the beam set as $B=\left\{b_{1}, b_{2}, \cdots, b_{N}\right\}$ with $|B|=N$, and the user set as $U=\left\{u_{1}, u_{2}, \cdots, u_{K}\right\}$ with $|U|=K$. Particularly, with $N_{l}=1$, $\forall l=1, 2, \cdots, L$, there is only one beam available at each AP. In this special case, the term ``beam'' can be used interchangeably with ``AP'' and the general cell-free wireless network considered in this paper reduces to a cell-free massive MIMO system that originates from the concepts of DAS and C-RAN. In another special case with $L=1$, there is only one AP in the network, that is, the network reduces to a single-cell fixed-beam based massive MIMO system. 

Since there are two types of nodes in the network, users and beams, and communications only happen between users and beams, let us represent the network by a weighted undirected bipartite graph $G=(V, E)$, where $V=U\cup B$ is the vertex set of users and beams, and $E=\left\{(u_{k}, b_{n}): \forall u_{k}\in U, \forall b_{n}\in B\right\}$ denotes the set of edges representing the beamspace channels between users and beams. The weight matrix of edges is denoted by $\mathbf{W}\in \mathbb{R}^{K\times N}$, where the $k$th row and $n$th column element $w_{k,n}$ is the edge weight between user $u_{k}$ and beam $b_{n}$. Let us rewrite the vertex set as $V=\left\{v_{1}, v_{2}, \cdots, v_{K+N}\right\}$ with the $i$th element $v_{i}=u_{i}$ if $1 \leq i\leq K$ and $v_{i}=b_{i-K}$ if $K+1 \leq i \leq K+N $. The weighted adjacency matrix $\mathbf{A}$ of graph $G$ is then given by
\begin{align}\label{A}
\mathbf{A}=\left[\begin{array}{*{20}{c}}
\mathbf{0}_{K\times K} & \mathbf{W} \\ 
\mathbf{W}^{T} &\mathbf{0}_{N\times N}
\end{array}\right].
\end{align}
%where $\mathbf{W}^{T}$ is the transpose of the weight matrix $\mathbf{W}$. 

\begin{figure}[t]
	\begin{center}
		\includegraphics[width=0.68\textwidth]{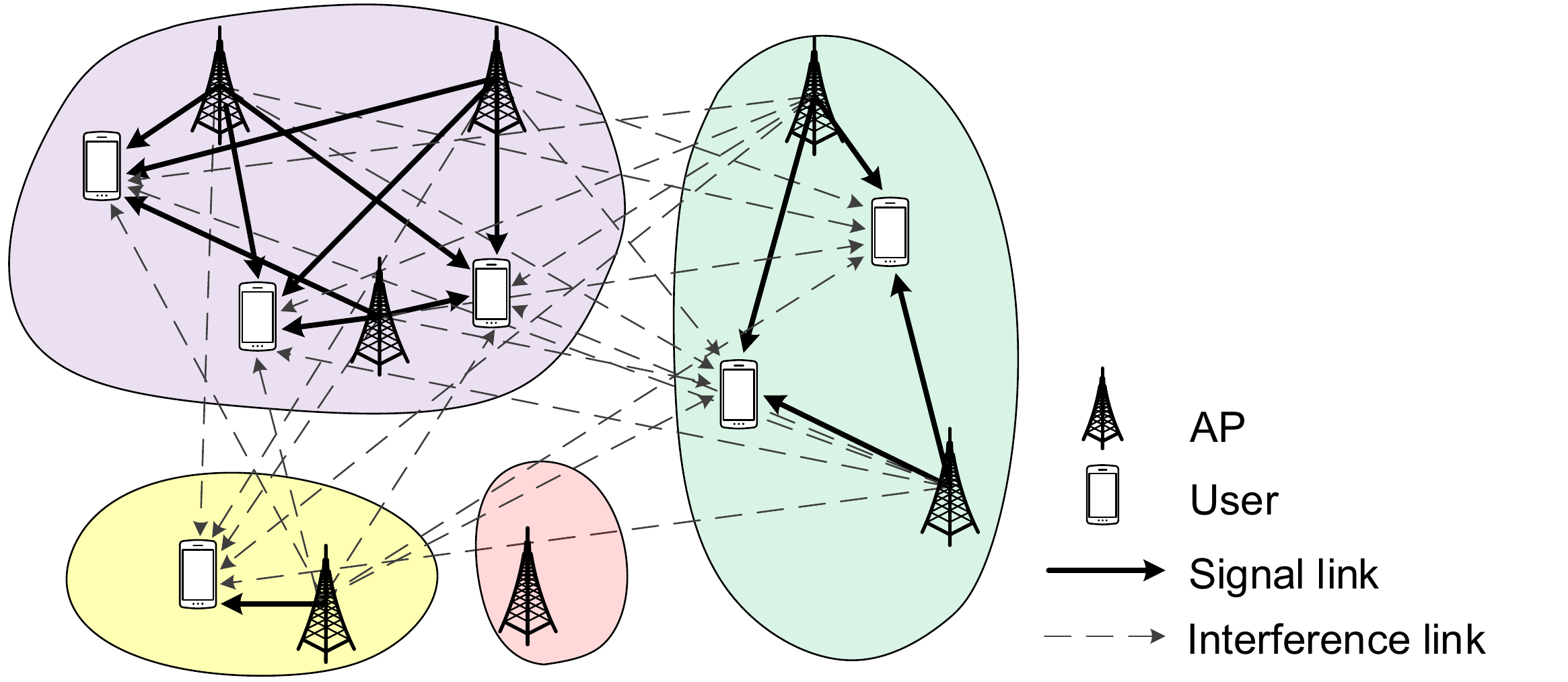}
		\caption{Illustration of clustered cell-free networking, where only one beam is assumed at each AP for simplicity.}
		\label{FIG_NetworkModel}
	\end{center}
\end{figure}

In a clustered cell-free network, as depicted in Fig. \ref{FIG_NetworkModel}, the whole network is decomposed into a number of small subnetworks that operate independently and in parallel, while joint processing can be performed in each subnetwork for intra-subnetwork interference cancellation. It should be noted that the beams available at one AP can be assigned to different subnetworks, which enables flexible cooperation among APs. If a subnetwork contains beams only, these beams will be turned off to save energy. If all the beams belonging to an AP are off, this AP will be switched into sleep mode. Let $M$ denote the number of decomposed subnetworks and $C_{m}$ denote the set of users and beams in the $m$th subnetwork. We have
\begin{align}\label{sum-Cm}
\bigcup_{m=1}^{M}C_{m}=U\cup B=V,
\end{align}
and 
\begin{align}\label{inter-Cm}
C_{m}\cap C_{m^{'}} = \emptyset,~\forall m^{'}\neq m.
\end{align}
It can be seen that by modeling the network as a graph, each subnetwork is a subgraph. The clustered cell-free networking problem can be then regarded as a graph partitioning problem. A partition $\mathcal{M}$ of graph $G$ with $M$ subgraphs is denoted as
\begin{align}\label{M-def}
\mathcal{M}=\left\{C_{1}, C_{2}, \cdots, C_{M}\right\}.
\end{align}
The main notations used in this paper are listed in Table 1 for ease of reading. 

\begin{table}
	\begin{center}
		\caption{Main Notations}
		\begin{tabular}{|c||c|}
			\hline
			$K$ & Number of users \\
			\hline
			$L$ & Number of APs \\
			\hline
			$N_l$ & Number of beams at AP $l$ \\
			\hline
			$N$ & Total number of beams \\
			\hline
			$B=\{b_{1}, b_{2}, \cdots, b_{N}\}$ & Beam set \\
			\hline
			$U=\{u_{1}, u_{2}, \cdots, u_{K}\}$ & User set \\
			\hline
			$G=(V, E)$ & A graph with vertex set $V$ and edge set $E$ \\
			\hline
			$\mathbf{W}$ & Edge weight matrix \\
			\hline
			$\mathbf{A}$ & Weighted adjacency matrix \\
			\hline
			$\mathbf{D}$ & Degree matrix \\
			\hline
			$\mathbf{L}$ & Graph Laplacian matrix \\
			\hline
			$M$ & Number of decomposed subnetworks \\
			\hline
 			$C_{m}$ & Set of users and beams in the $m$th subnetwork \\
 			\hline
 			$\mathcal{M}=\{C_{1}, C_{2}, \cdots, C_{M}\}$ & A graph partition with $M$ subgraphs \\
 			\hline
 			Cut($C_m$) & Cut function of subgraph $C_{m}$ \\
 			\hline
 			$R(\mathcal{M})$ & Per-user rate achieved under partition $\mathcal{M}$ \\
 			\hline
 			$R_{th}$ & Minimum per-user rate requirement \\
 			\hline
 			$\mu_{k}$ & Received signal-to-interference-plus-noise ratio (SINR) at user $u_{k}$ \\
 			\hline
 			$P_{t}$ & Transmit power on each beam \\
 			\hline
 			$\sigma^2$ &Variance of additive white Gaussian noise (AWGN) \\
 			\hline
 			$g_{k,n}$ & Beamspace channel gain coefficient from beam $b_{n}$ to user $u_{k}$ \\
 			\hline
 			$b_{n_{k}^{*}}$ & Best beam for user $u_{k}$ with the highest beamspace channel gain \\
 			\hline
			$\alpha$ & Path-loss exponent \\
			\hline
		\end{tabular}
	\end{center}
\end{table}

\subsection{Problem Formulation of Rate-Constrained Clustered Cell-Free Networking}
%Note that a user's rate performance improves as its subnetwork size increases thanks to the joint transmission gain yet at the cost of increased complexity and signaling overhead. To minimize the system complexity and signaling overhead as much as  possible, in this paper, we aim to maximize the number of decomposed subnetworks under a per-user rate constraint.

In this paper, it is assumed that the intra-subnetwork interference is canceled by some efficient precoding scheme in each subnetwork. To perform such joint processing, the beamspace CSI between beams and users in each subnetwork is required. Therefore, the subnetwork size determines the signaling overhead and complexity of joint processing. With more subnetworks, the joint processing complexity and signaling overhead of each subnetwork is reduced on average. Meanwhile, the rate performance degrades due to the increased inter-subnetwork interference. Under the constraint that the per-user rate must be no smaller than a given threshold $R_{th}$ for service quality guarantee, this paper aims to maximize the number of decomposed subnetworks $M$ as much as   possible, so as to reduce the joint processing complexity and signaling overhead to the most extent. The clustered cell-free networking problem can be formulated as
\begin{align}\label{def-Prob}
\mathcal{P}1:~\max_{\mathcal{M}=\left\{C_{1}, C_{2}, \cdots, C_{M}\right\}}\quad &M \\ 
\text{s.t.}\qquad \quad &R(\mathcal{M}) \geq R_{th}, \label{const-minR} \\
&\bigcup_{m=1}^{M}C_{m}=U\cup B, \label{const-sumCm}\\
&C_{m}\cap C_{m^{'}} = \emptyset,~\forall m^{'}\neq m \label{const-inter-Cm}, \\
&C_{m}\cap B \neq \emptyset,~\forall m, \label{const-iso-user}
\end{align}
where (\ref{const-sumCm}) and (\ref{const-inter-Cm}) follow the constraints given in (\ref{sum-Cm}) and (\ref{inter-Cm}).  (\ref{const-iso-user}) ensures that there is at least one beam in each subnetwork such that all the users in the system can be served. $R(\mathcal{M})$ is the per-user rate achieved under partition $\mathcal{M}$, which is defined as
\begin{align}\label{def-AveR}
	R(\mathcal{M}) \triangleq  \frac{1}{K}\sum_{m=1}^{M}\sum_{u_{k}\in C_{m}}R_{k},
	%=\frac{1}{K}\sum_{m=1}^{M}\sum_{u_{k}\in C_{m}}\log_{2}\left(1+\mu_{k}\right).
\end{align}
where $R_{k}$ is the achievable rate of user $u_{k}$. By normalizing the total system bandwidth into unity and focusing on the spectral efficiency, the achievable rate $R_{k}$ of user $u_{k}$ can be expressed as
\begin{align}\label{def-Rk}
	R_{k}=\log_{2}\left(1+\mu_{k}\right),
\end{align}
where $\mu_{k}$ is the received signal-to-interference-plus-noise ratio (SINR) at user $u_{k}$.

Note that the received SINR $\mu_{k}$ is closely dependent on the precoding scheme adopted in each subnetwork. In this paper, instead of focusing on a specific precoding scheme, which may limit the application scenarios of the proposed clustered cell-free networking algorithm, we adopt an approximation of the received SINR $\mu_{k}$ for a general spatial precoding scheme with a large number of beams and/or APs. Specifically, with a large number of geographically distributed APs and/or preformed narrow and directive beams available at each AP, the beamspace channel of a user is highly sparse, i.e., the total channel gain of a user is mainly contributed by its dominant component\cite{Junyuan_BeamPerformance,Junyuan_DAS}. As a result, the SINR of reference user $u_{k}\in C_{m}$ in the $m$th subnetwork can be approximated by
\begin{align}\label{def-mu}
	\mu_{k}\mathop{\approx}\frac{P_{t}|g_{k,n_{k}^{*}}|^{2}}{\sum_{b_{n}\notin C_{m}, b_{n}\in B}P_{t}|g_{k,n}|^{2}+\sigma^2},
\end{align}
where $g_{k,n}$ denotes the beamspace channel gain coefficient from beam $b_{n}$ to user $u_{k}$. $b_{n_{k}^{*}}$ is the best beam for user $u_{k}$ with the highest beamspace channel gain, i.e., $n_{k}^{*}=\arg \max_{n=1, 2, \cdots, N} |g_{k, n}|^2$. $P_{t}$ is the transmit power on each beam. $\sigma^{2}$ is the variance of the additive white Gaussian noise (AWGN). Note that the SINR approximation in (\ref{def-mu}) was shown to be accurate when zero-forcing dirty paper coding (ZF-DPC) is adopted to mitigate the intra-subnetwork interference \cite{Veetil}. In general, (\ref{def-mu}) serves as a good approximation given that the intra-subnetwork interference is eliminated by some spatial precoding scheme\cite{Junyuan_DAS,Liu,Veetil}.

\section{Rate-Constrained Clustered Cell-Free Networking}
As the formulated optimization problem $\mathcal{P}1$ is difficult to solve due to the mixture of the continuous constraint in (\ref{const-minR}) and the combinatorial constraints in (\ref{const-sumCm})--(\ref{const-iso-user}), in this section, we will first transform problem $\mathcal{P}1$ into a cut-constrained bipartite graph partitioning problem, and then propose a rate-constrained network decomposition algorithm for clustered cell-free networking.

\subsection{Problem Reformulation}
Since $f(x)=\log_{2}(1+1/x)$ is a convex function for $x>0$, a lower-bound of the per-user rate, $R^{lb}(\mathcal{M})$, can be obtained from (\ref{def-AveR})--(\ref{def-mu}) as
\begin{align}\label{AveR-lb}
	R(\mathcal{M})\geq R^{lb}(\mathcal{M})=
		\log_{2}\left(1+\frac{K}{\beta+\sum_{m=1}^{M}\sum_{u_{k}\in C_{m}}\sum_{b_n\notin C_{m}, b_{n}\in B}\frac{|g_{k,n}|^2}{|g_{k, n_{k}^{*}}|^2}}\right),
\end{align}
according to Jensen's inequality, where $\beta=\sum_{k=1}^{K}\frac{\sigma^2}{P_{t}|g_{k, n_{k}^{*}}|^2}$.

Note that the cut function $\text{cut}(C_{m})$ of a subgraph $C_{m}$ is defined as the sum weights of the edges between the vertices inside and outside $C_{m}$, given by
\begin{align}\label{def-cut}
\text{cut}(C_{m})=\sum_{v_{i}\in C_{m}} \sum_{v_{j}\in \overline{C}_{m}} a_{i,j},
\end{align}
where $a_{i,j}$ denotes the $i$th row and $j$th column element of adjacency matrix $\mathbf{A}$, and $\overline{C}_{m}=V\setminus C_{m}$ is the complement of subset $C_{m}\subseteq V$. According to (\ref{A}), cut($C_{m}$) can be rewritten as
\begin{align}\label{cut}
\text{cut}(C_{m})=\sum_{u_{k}\in C_{m}}\sum_{b_{n}\in \overline{C}_{m}} w_{k,n}+\sum_{b_{n}\in C_{m}}\sum_{u_{k}\in \overline{C}_{m}} w_{k,n}.
\end{align}
Due to the symmetry of adjacency matrix $\mathbf{A}$, the sumcut function of all the $M$ subgraphs that form a partition of graph $G$ can be obtained as
\begin{align}\label{sum-cut}
\sum_{m=1}^{M}\text{cut}(C_{m})
%=2\sum_{m=1}^{M}	\sum_{u_{k}\in C_{m}}\sum_{b_{n}\in \overline{C}_{m}} w_{k,n}
=2\sum_{m=1}^{M}\sum_{u_{k}\in C_{m}} \sum_{b_{n}\in B, b_n\notin C_{m}} w_{k,n}.
\end{align} 

Let us model the edge weight $w_{k,n}\in \mathbf{W}$ between user $u_{k}$ and beam $b_{n}$ as
\begin{align}\label{def-w}
w_{k,n}=\frac{|g_{k,n}|^2}{|g_{k,n_{k}^{*}}|^2}.
\end{align}
By combining (\ref{AveR-lb}) and (\ref{sum-cut})--(\ref{def-w}), the per-user rate lower-bound $R^{lb}(\mathcal{M})$ can be rewritten as
\begin{align}\label{AveR-lb-cut}
%R(\mathcal{M})\geq 
R^{lb}(\mathcal{M})=
\log_{2}\left(1+\frac{K}{\beta+\frac{1}{2}\sum_{m=1}^{M}\text{cut}(C_{m})}\right).
\end{align}
With $R^{lb}(\mathcal{M})\geq R_{th}$, the per-user rate constraint in (\ref{const-minR}) is always satisfied. Therefore, by replacing $R(\mathcal{M})$ with its lower-bound $R^{lb}(\mathcal{M})$, the clustered cell-free networking problem $\mathcal{P}1$  given in (\ref{def-Prob})--(\ref{const-iso-user}) can be transformed to 
%and combine (\ref{AveR-lb})--(\ref{const-iso-user-lb}) and (\ref{sum-cut})--(\ref{def-w}), the network decomposition problem can be rewritten into the following form
\begin{align}\label{Prob-sumcut}
\mathcal{P}2:~\max_{\mathcal{M}=\left\{C_{1}, C_{2}, \cdots, C_{M}\right\}}\quad &M \\ 
\text{s.t.} \quad \qquad &\sum_{m=1}^{M}\text{cut}(C_{m})\leq \frac{2K}{2^{R_{th}}-1}-2\beta, \label{const-sumcut} \\
&\bigcup_{m=1}^{M}C_{m}=U\cup B, \label{const-sumCm-sumcut}\\
&C_{m}\cap C_{m^{'}} = \emptyset,~\forall m^{'}\neq m \label{const-inter-Cm-sumcut}, \\
&C_{m}\cap B \neq \emptyset,~\forall m. \label{const-iso-user-sumcut}
\end{align}

\textbf{Remark:} Different from \cite{Lin} where APs are always activated, the clustered cell-free networking problem formulated in this paper could not only produce balanced subnetworks to reduce system complexity and signaling overhead efficiently, but also switch off beams that are badly aligned with users or associated with APs far from users to save energy. Specifically, by defining the edge weight between a user and a beam as the normalized beamspace channel gain, a beam with no users close to its main direction or its associated AP would have small edge weights, and thus lead to small sumcut when forming a subnetwork alone. With the aim of maximizing the total number of subnetworks, isolating such beams would be prioritized, which facilitates the on-off switch of beams for energy saving. From the perspective of users, the maximum edge weight incident to each user is always 1, indicating balanced connectivity of users on the constructed graph $G$, and thus balanced sizes of the subnetworks containing both users and beams.

\subsection{Rate-Constrained Network Decomposition Algorithm }
To solve the clustered cell-free networking problem $\mathcal{P}2$ given in (\ref{Prob-sumcut})--(\ref{const-iso-user-sumcut}), let us take the double-loop approach in \cite{Lin}. Specifically, in the inner loop, an $M$-way graph partitioning problem is solved to find the optimal partition $\mathcal{M}_{|M}^{*}$ that minimizes the sumcut $\sum_{m=1}^{M} \text{cut}(C_{m|M})$ given the number of subgraphs $M$. In the outer loop, the maximum number of subgraphs $M^{*}$ is searched as long as its corresponding $M$-way partition $\mathcal{M}_{|M^{*}}^{*}$ satisfies the sumcut constraint in (\ref{const-sumcut}). 

\subsubsection{$M$-Way Partitioning}
In the inner loop, we aim at finding the optimal partition $\mathcal{M}_{|M}^{*}=\left\{C_{1|M}^{*}, C_{2|M}^{*}, \cdots, C_{M|M}^{*}\right\}$ to minimize the sumcut, which can be presented as
\begin{align}\label{Prob-1-cut}
\mathcal{P}3:~\min_{\mathcal{M}_{|M}=\left\{C_{1|M}, C_{2|M}, \cdots, C_{M|M}\right\}}\quad &\sum_{m=1}^{M}\text{cut}(C_{m|M}) \\ 
\text{s.t.} \qquad \qquad \quad &\bigcup_{m=1}^{M}C_{m|M}=U\cup B, \label{const-sumCm-sub1-cut}\\
&C_{m|M}\cap C_{m^{'}|M} = \emptyset,~\forall m^{'}\neq m \label{const-inter-Cm-sub1-cut}, \\
&C_{m|M}\cap B \neq \emptyset,~\forall m. \label{const-iso-user-sub1-cut}
\end{align}
In order to solve problem $\mathcal{P}3$, let us first convert the bipartite graph partitioning problem into a graph partitioning problem. 

\begin{figure}[t]
	\begin{center}
		\includegraphics[width=1\textwidth]{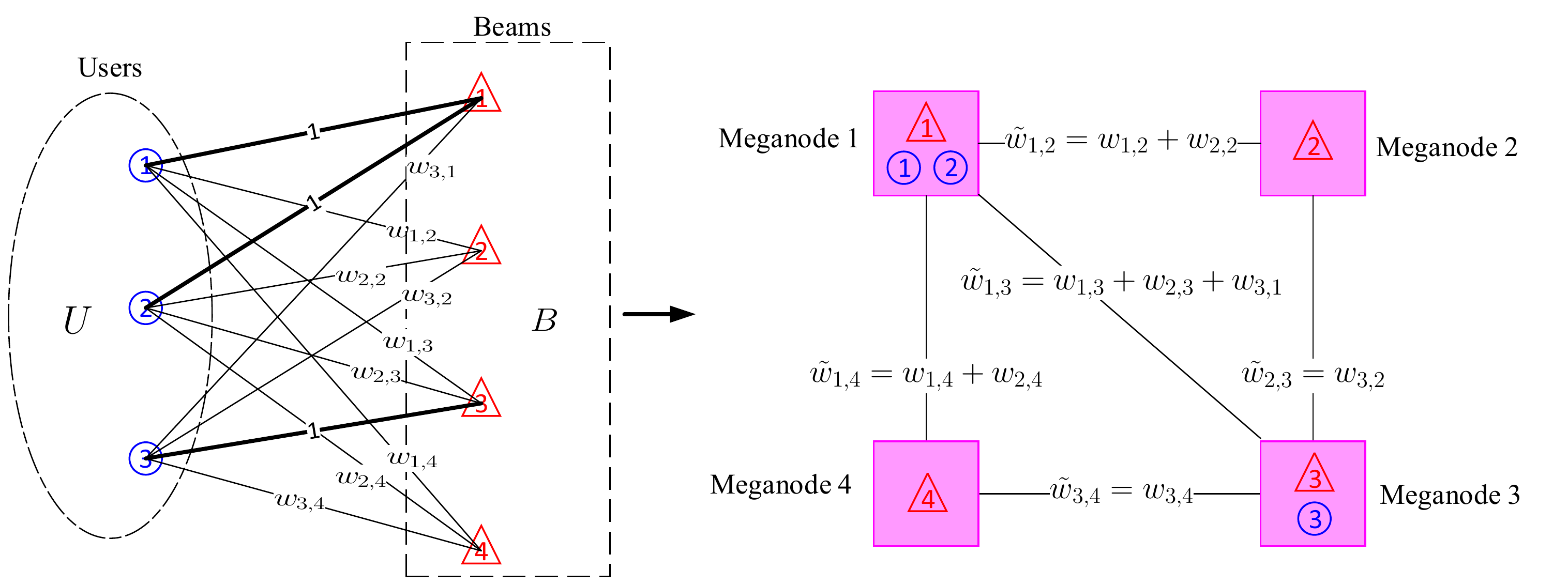}
		\caption{A toy example of 3 users and 4 beams to illustrate the construction of graph $\tilde{G}=(\tilde{V}, \tilde{E})$. In the bipartite graph on the left-hand side, the edges with weight 1 are plotted in thick lines.}
		\label{FIG_toy_example}
	\end{center}
\end{figure}

By merging each beam vertex $b_{i}$ with the user vertices with edge weight of 1 between them to form a new meganode $\tilde{v}_{i}$ as 
\begin{align}\label{meganode}
	\tilde{v}_{i}=\left\{b_{i}\cup \{u_{k}:n_{k}^{*}=b_{i}, \forall u_{k}\in U\}\right\}, 
\end{align}
a new graph $\tilde{G}=(\tilde{\mathcal{V}}, \tilde{E})$ of $N$ meganodes can be constructed, where $\tilde{\mathcal{V}}=\{\tilde{v}_{1}, \tilde{v}_{2}, \cdots, \tilde{v}_{N}\}$ is the meganode set. $\tilde{E}=\left\{(\tilde{v}_{i}, \tilde{v}_{j}): \forall \tilde{v}_{i}\in \tilde{\mathcal{V}}, \forall \tilde{v}_{j}\in \tilde{\mathcal{V}}, i\neq j\right\}$ is the set of edges, where the edge weight $\tilde{w}_{i,j}$ between meganode $\tilde{v}_{i}$ and meganode $\tilde{v}_{j}$ is defined as the sum weight of the edges in the original bipartite graph $G$ connecting the vertices inside meganode $\tilde{v}_{i}$ and the vertices inside meganode $\tilde{v}_{j}$, that is,
\begin{align}\label{edge-new}
\tilde{w}_{i,j}=\left\{\begin{array}{*{20}{l}}
\sum_{u_{k}\in \tilde{v}_{i}} w_{k,j}+\sum_{u_{k}\in \tilde{v}_{j}} w_{k,i}, & \text{if~}i \neq j; \\
0, & \text{otherwise}.
\end{array}\right.
\end{align}
A toy example is presented in Fig. \ref{FIG_toy_example} to illustrate the construction of the new graph $\tilde{G}=(\tilde{\mathcal{V}}, \tilde{E})$, where beam 1 is merged with users 1 and 2 to form meganode 1, and beam 3 and user 3 are merged as meganode 3. 
Appendix A shows that the optimization problem $\mathcal{P}3$ given in (\ref{Prob-1-cut})--(\ref{const-iso-user-sub1-cut}) is equivalent to the following graph partitioning problem on graph $\tilde{G}=(\tilde{\mathcal{V}}, \tilde{E})$:
\begin{align}\label{Prob-1-new}
\mathcal{P}4:~\min_{\mathcal{\tilde{M}}_{|M}=\left\{\tilde{\mathcal{C}}_{1|M}, \tilde{\mathcal{C}}_{2|M}, \cdots, \tilde{\mathcal{C}}_{M|M}\right\}}\quad &\sum_{m=1}^{M} \text{cut}(\tilde{\mathcal{C}}_{m|M}) \\ 
\text{s.t.} \quad \qquad \qquad &\bigcup_{m=1}^{M}\tilde{\mathcal{C}}_{m|M}=\tilde{\mathcal{V}}, \label{const-sumCm-sub1-new}\\
&\tilde{\mathcal{C}}_{m|M}\cap \tilde{\mathcal{C}}_{m^{'}|M} = \emptyset,~\forall m^{'}\neq m, \label{const-inter-Cm-sub1-new}
\end{align}
where $\tilde{\mathcal{M}}_{|M}=\left\{\tilde{\mathcal{C}}_{1|M}, \tilde{\mathcal{C}}_{2|M}, \cdots, \tilde{\mathcal{C}}_{M|M}\right\}$ denotes a partition of graph $\tilde{G}$ with $M$ subgraphs. 

It can be seen that the optimization problem $\mathcal{P}4$ is a canonical mincut problem, which can be efficiently solved by a spectral clustering method\cite{Luxburg}. Specifically, let
\begin{align}\label{L}
\mathbf{L}=\mathbf{D}-\mathcal{\mathbf{\tilde{W}}}
\end{align}
denote the graph Laplacian matrix of graph $\tilde{G}$, where $\tilde{\mathbf{W}}$ is the weighted adjacency matrix with each element $\tilde{w}_{i,j}$ defined in (\ref{edge-new}). $\mathbf{D}$ is the degree matrix given by
\begin{align}\label{D}
\mathbf{D}=diag(d_{1}, d_{2}, \cdots, d_{N}),
\end{align}
where $d_{i}$ is the degree of meganode $\tilde{v}_{i}$ with
\begin{align}\label{di}
d_{i}=\sum_{j=1}^{N}\tilde{w}_{i,j}.
\end{align}
By computing the eigenvectors $\mathbf{p}_{1}, \mathbf{p}_{2}, \cdots, \mathbf{p}_{M}$ of graph Laplacian matrix $\mathbf{L}$ corresponding to the $M$ smallest eigenvalues with $\mathbf{p}_{m}\in\mathbb{R}^{N\times 1}$, $m=1, 2, \cdots, M$, a new matrix
\begin{align}\label{Y}
\mathbf{Y}=[\mathbf{p}_{1}, \mathbf{p}_{2}, \cdots, \mathbf{p}_{M}]
\end{align}
can be formed. Let $\mathbf{y}_{i}^{r} \in \mathbb{R}^{1\times M}$ be the $i$th row vector of $\mathbf{Y}$. By clustering the $N$ vectors $\mathbf{y}_{1}^{r},\mathbf{y}_{2}^{r}, \cdots, \mathbf{y}_{N}^{r}$ into $M$ clusters $\hat{C}_{1}, \hat{C}_{2}, \cdots, \hat{C}_{M}$ with the $k$-means algorithm, the optimal $\tilde{\mathcal{C}}_{m|M}^{*}$ for any $m=1,2, \cdots, M$ can be obtained as
\begin{align}\label{opt-Cm}
\tilde{\mathcal{C}}_{m|M}^{*}=\left\{\tilde{v}_{i}: \mathbf{y}_{i}^{r} \in \hat{C}_{m}, i=1, 2, \cdots, N\right\}.
\end{align}
By combining (\ref{meganode}) and (\ref{opt-Cm}), the optimal partition $\mathcal{M}_{|M}^{*}=\left\{C_{1|M}^{*}, C_{2|M}^{*}, \cdots, C_{M|M}^{*}\right\}$ can be obtained with
\begin{align}\label{opt-Cm-final}
C_{m|M}^{*}=\left\{b_{i}: \tilde{v}_{i} \in \tilde{\mathcal{C}}_{m|M}^{*}\right\} \cup \left\{u_{k}: u_{k}\in \tilde{v}_{i}, \tilde{v}_{i} \in \tilde{\mathcal{C}}_{m|M}^{*} \right\}, m=1, 2, \cdots, M.
\end{align}

\subsubsection{Maximization of $M$}
Based on the solution of the inner-loop $M$-way partitioning problem, $\mathcal{M}_{|M}^{*}=\left\{C_{1|M}^{*}, C_{2|M}^{*}, \cdots, C_{M|M}^{*}\right\}$, the outer-loop problem can be written as
\begin{align}\label{Prob-2}
\mathcal{P}5:~\max \quad &M \\ \label{const-2}
\text{s.t.} \quad &\sum_{m=1}^{M} \text{cut}(C_{m|M}^{*}) \leq \frac{2K}{2^{R_{th}}-1}-2\beta.
\end{align}

As the exhaustive search of the optimal number of decomposed subnetworks $M^{*}$ would lead to high computational complexity with a massive number of beams $N$, based on the monotonic property of the mincut function $\sum_{m=1}^{M}\text{cut}(C_{m|M}^{*})$ shown in Theorem 1, $M^{*}$ can be obtained by adopting a binary search approach. 
\begin{theorem}
	The mincut function $\sum_{m=1}^{M}\text{cut}(C_{m|M}^{*})$ satisfies 
	\begin{align}\label{mincut-monotone}
	\sum_{m=1}^{M}\text{cut}(C_{m|M}^{*}) \leq \sum_{m=1}^{M+1}\text{cut}(C_{m|M+1}^{*}),
	\end{align}
	for any $M=1, 2, \cdots, N-1$.
\end{theorem} 
\begin{IEEEproof}
	See Appendix A. 
\end{IEEEproof}

\begin{algorithm}[t]
	\caption{RC-NetDecomp}
	\label{Alg}
	\begin{algorithmic}[1]
		\State \textbf{Input:}  Minimum per-user rate requirement $R_{th}$, weight matrix $\mathbf{W}$.
		\State \textbf{Initialization:} $min=1$, $max=N+1$, $\mathcal{M}^{*}=\left\{V\right\}$.
		\While{$min<max-1$}
		\State $M=\lfloor\frac{min+max}{2}\rfloor$;
		\State Compute $\tilde{\mathbf{W}}$, $\mathbf{L}$ and $\mathbf{Y}$ based on (\ref{edge-new}) and (\ref{L})--(\ref{Y}); 
		\State Run the $k$-means algorithm to cluster row vectors of $\mathbf{Y}$, $\mathbf{y}_{1}^{r}, \mathbf{y}_{2}^{r}, \cdots, \mathbf{y}_{N}^{c}$, into $M$ clusters $\hat{C}_{1}, \hat{C}_{2}, \cdots, \hat{C}_{M}$;
		\State Compute $\mathcal{M}_{|M}^{*}=\left\{C_{1|M}^{*}, C_{2|M}^{*}, \cdots, C_{M|M}^{*}\right\}$ according to (\ref{opt-Cm}) and (\ref{opt-Cm-final});
		\If{$\sum_{m=1}^{M}\text{cut}(C_{m|M}^{*})\leq \frac{2K}{2^{R_{th}}-1}-2\beta$}
		\State $min=M$, $\mathcal{M}^{*}=\mathcal{M}_{|M}^{*}$;
		\Else 
		\State $max=M$;
		\EndIf
		\EndWhile
		%\State $M^{*}=min$.
		\State \textbf{Output:} $\mathcal{M}^{*}$.
	\end{algorithmic}
\end{algorithm}

By combining the spectral clustering algorithm solving the $M$-way partitioning subproblem in the inner loop and the binary search algorithm solving the maximization of $M$ subproblem in the outer loop, a rate-constrained network decomposition (RC-NetDecomp) algorithm is proposed for clustered cell-free networking with detailed steps summarized in Algorithm 1. 

\subsection{Complexity Analysis}
Since a binary search method is adopted in the outer loop to find the maximum number of subnetworks $M^{*}$ on graph $\tilde{G}$, where the number of vertices is $N$, the computational complexity of the outer-loop binary search is $O(\log N)$. For the inner-loop that partitions graph $\tilde{G}$ into $M$ subgraphs by adopting a spectral clustering method, the complexity is mainly determined by the eigenvalue decomposition of the Laplacian matrix $\mathbf{L}$ of graph $\tilde{G}$, which has the computational complexity of $O(N^{3})$. The overall computational complexity of the proposed RC-NetDecomp algorithm is then $O(N^{3}\log N)$. Note that by constructing the new graph $\tilde{G}$, the number of vertices in the graph partitioning problem reduces from $K+N$ to $N$, leading to lower computational complexity than partitioning the original bipartite graph $G$ directly as a byproduct.

\section{Simulation Results}
In this section, simulation results are presented to demonstrate the performance of the proposed RC-NetDecomp algorithm for clustered cell-free networking. A cell-free wireless network is considered, where users and APs are assumed to be randomly distributed by following an independent and identical uniform distribution. Note that decomposing a network according to instantaneous channel gains leads to rapidly varying network structure, which is difficult to be applied in practical systems. To this end, let us only consider the path-loss and beamforming gain when performing the RC-NetDecomp algorithm. The beamspace channel gain from beam $b_{n}$ to user $u_{k}$ is then set as $|g_{k,n}|^2=d_{k,n}^{-\alpha}\cdot D_{k,n}$, where $d_{k,n}$ is the distance between user $u_{k}$ and the AP to which beam $b_{n}$ belongs, and $\alpha$ is the path-loss exponent. $D_{k,n}$ denotes the power gain of beam $b_{n}$ at user $u_{k}$. The corresponding edge weight $w_{k,n}$ given in (\ref{def-w}) can be then written as 
\begin{align}\label{w}
	w_{k,n}=\frac{d_{k,n}^{-\alpha}\cdot D_{k,n}}{d_{k,n_{k}^{*}}^{-\alpha}\cdot D_{k,n_{k}^{*}}},
\end{align}
where $n_{k}^{*}=\arg \max_{n=1, 2, \cdots, N} d_{k,n}^{-\alpha} D_{k,n}$. After performing the RC-NetDecomp algorithm, zero-forcing beamforming (ZFBF) is applied in each subnetwork. Since ZFBF requires that the number of beams is no less than the number of users, the rate of each user is zero if there are more users than beams in a subnetwork.

As discussed in Section I, there are two kinds of algorithms in the literature that could be deployed for clustered cell-free networking.
\begin{itemize}
	\item User-centric clustering algorithms: Subnetworks are formed by merging user-centric virtual cells. A representative user-centric clustering algorithm was proposed in \cite{Junyuan_VC}, where each user first chooses $S$ APs with the highest channel gains, and then the overlapped virtual cells are merged. 
	\item AP-centric clustering algorithms: APs are first clustered and then each user joins the subnetwork containing its associated AP. A representative AP-centric clustering algorithm was proposed in \cite{Goldsmith}, where a hierarchical clustering algorithm with minimax linkage criterion was adopted to cluster APs.
\end{itemize}

To demonstrate the performance gains of the proposed RC-NetDecomp algorithm, the AP-centric algorithm in \cite{Goldsmith} and the user-centric algorithm in \cite{Junyuan_VC} are adopted as benchmarks. Note that the RC-NetDecomp algorithm is proposed for a general cell-free wireless network with $L$ APs and $N_{l}$ beams available at AP $l$, which reduces to the special cases of cell-free massive MIMO with $N_{l}=1$, $\forall l=1, 2, \cdots, L$, and single-cell massive MIMO with $L=1$. In the following, we will compare the proposed RC-NetDecomp algorithm against the aforementioned two baselines in these two special cases. Performance metrics of interest include the average per-user rate $\bar{R}\triangleq\mathbb{E}_{\{|g_{k,n}|^{2}\}_{u_{k}\in U, b_{n}\in B}}[R]$, the average minimum rate of users $\bar{R}_{min}\triangleq\mathbb{E}_{\{|g_{k,n}|^{2}\}_{u_{k}\in U, b_{n}\in B}}[\min_{u_{k}\in U}R_{k}]$, the average variance of users' rates $\bar{R}_{var}\triangleq\mathbb{E}_{\{|g_{k,n}|^{2}\}_{u_{k}\in U, b_{n}\in B}}\left[\tfrac{1}{K}\sum_{u_{k}\in U} (R_{k}-R)^{2}\right]$, the joint processing complexity indicated by the average number of decomposed subnetworks $\bar{M}^{*}\triangleq\mathbb{E}_{\{|g_{k,n}|^{2}\}_{u_{k}\in U, b_{n}\in B}}[M^{*}]$ along with the average maximum subnetwork size  $\overline{|C|}_{max}\triangleq\mathbb{E}_{\{|g_{k,n}|^{2}\}_{u_{k}\in U, b_{n}\in B}}\left[\max_{m=1, 2, \cdots, M^{*}}|C_{m|M^{*}}^{*}|\right]$, and the system energy efficiency indicated by the average percentage of beams switched off $\bar{P}_{off}\triangleq\mathbb{E}_{\{|g_{k,n}|^{2}\}_{u_{k}\in U, b_{n}\in B}}\left[\tfrac{1}{N}\sum_{m=1, C_{m|M^{*}}^{*}\cap U=\emptyset }^{M^{*}}|C_{m|M^{*}}^{*}|\right]$, all of which are obtained by averaging over 500 random realizations of users' and APs' locations.  

\vspace{-3mm}
\subsection{Special Case of Cell-Free Massive MIMO with $N_{l}=1$}
With the number of beams $N_{l}=1$ for any $l=1, 2, \cdots, L$, each AP is equipped with a single omnidirectional antenna and the general cell-free wireless network considered in this paper reduces to a cell-free massive MIMO system. In such a special case, a beam is equivalent to an AP, and there is no beamforming gain, i.e., $D_{k,n}=1$, $\forall b_{n}\in B, \forall u_{k} \in U$. As a result, the edge weight in (\ref{w}) reduces to $w_{k,n}=d_{k,n}^{-\alpha}/d_{k,n_{k}^{*}}^{-\alpha}$,
%\begin{align}\label{w-d}
%	w_{k,n}=\left(\frac{d_{k,n}}{d_{k,n_{k}^{*}}}\right)^{-\alpha},
%\end{align}
where $d_{k,n_{k}^{*}}$ equals the smallest distance from user $u_{k}$ to the $L$ APs, i.e., $d_{k,n_{k}^{*}}=\min_{n=1, 2, \cdots, L} d_{k,n}$.

\begin{figure*}[t]
	\centering
	{	\subfloat[$R_{th}=3.5$ bit/s/Hz]{\includegraphics[width=0.335\textwidth]{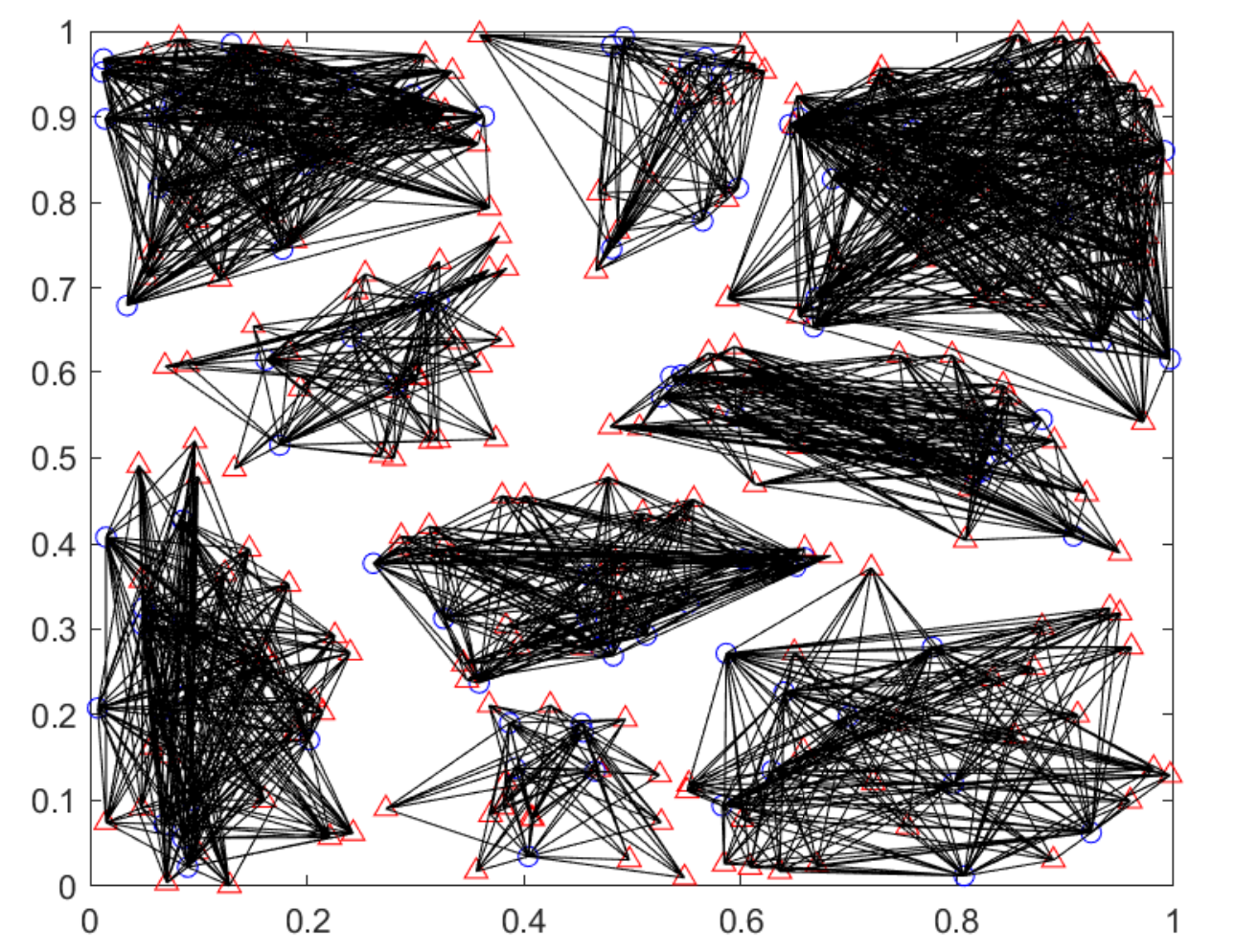}
			\label{FIG_Snapshot_Rmin3dot5}}
		\subfloat[$R_{th}=2$ bit/s/Hz]{\includegraphics[width=0.335\textwidth]{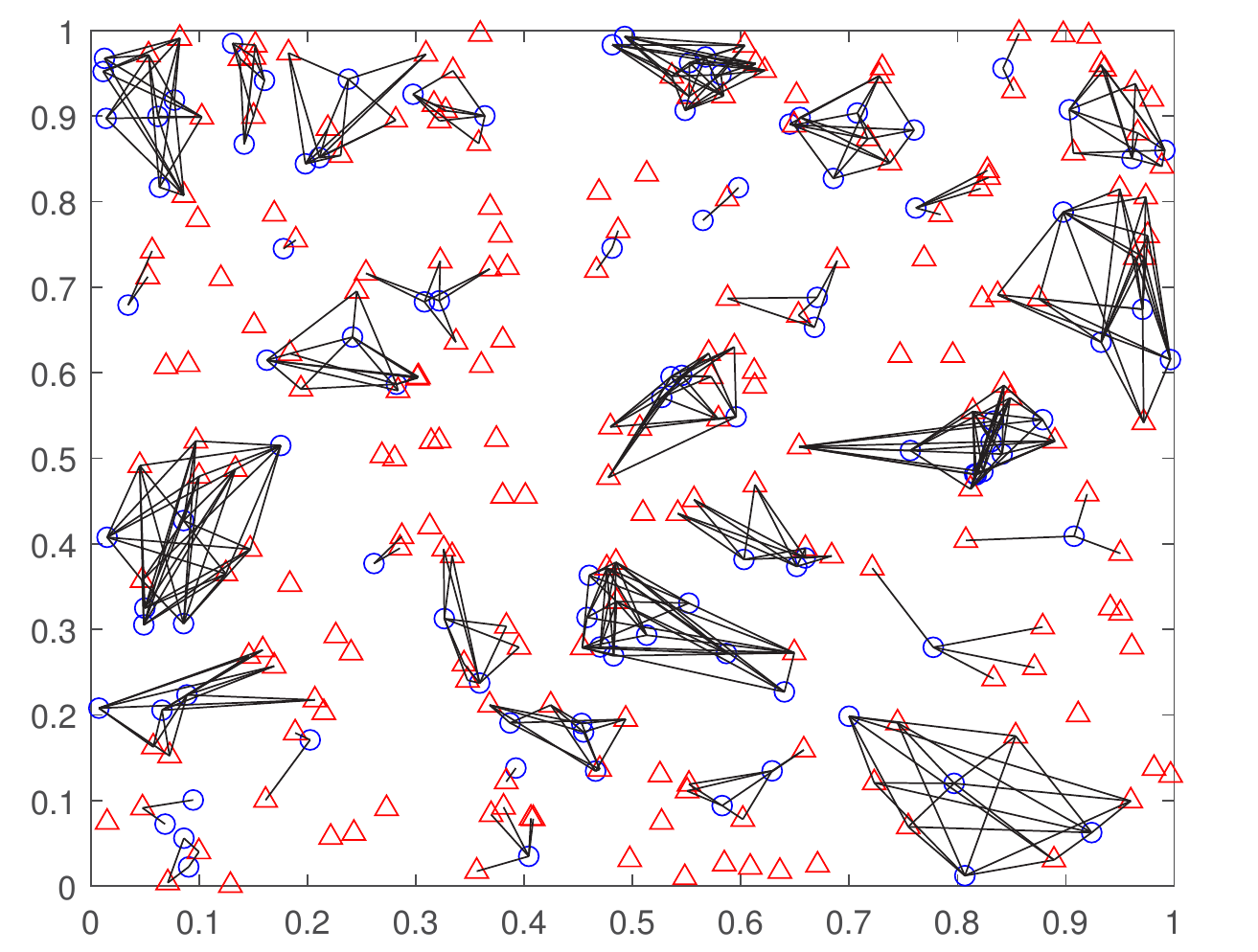}
			\label{FIG_Snapshot_Rmin2}} 
		\subfloat[$R_{th}=0$]{\includegraphics[width=0.335\textwidth]{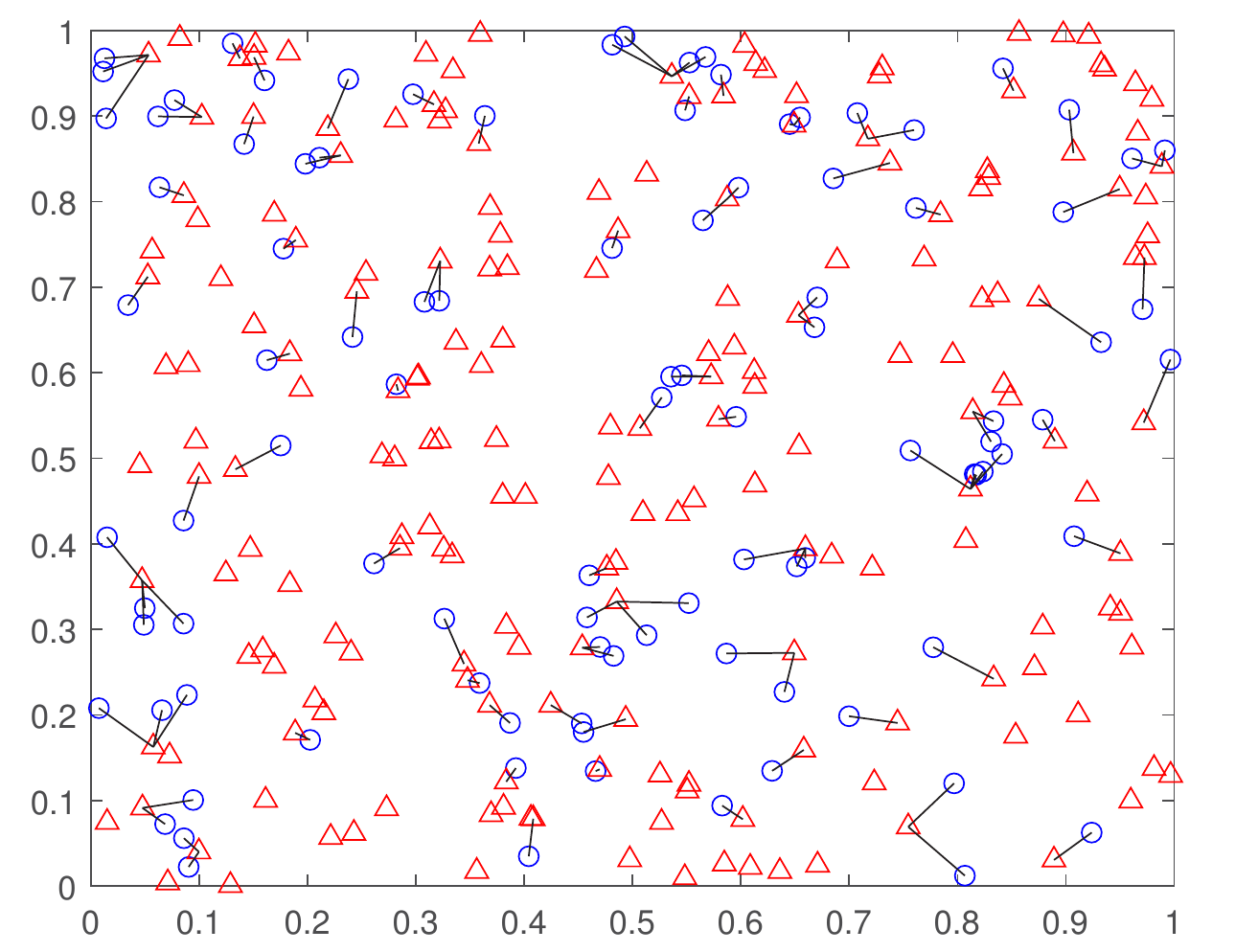}
			\label{FIG_Snapshot_Rmin0}}
	}
	\caption{Decomposed subnetworks of a randomly generated cell-free massive MIMO system by adopting the proposed RC-NetDecomp algorithm under various required per-user rate $R_{th}$. ``$\textcolor{blue}{\circ}$'' represents a user and ``$\textcolor{red}{\triangle}$'' represents an AP. $L=200$, $K=100$, $\alpha=4$.}
	\label{FIG_Snapshot_DAS}
\end{figure*}

\subsubsection{Decomposition Results}
Fig. \ref{FIG_Snapshot_DAS} illustrates the decomposition results of a cell-free massive MIMO system with a randomly generated topology by adopting the proposed RC-NetDecomp algorithm under various values of the per-user rate constraint $R_{th}$. It can be seen in Fig. \ref{FIG_Snapshot_Rmin3dot5} that with a relatively high $R_{th}=3.5$ bit/s/Hz, the whole network is decomposed into 9 subnetworks, and all the APs are involved in joint  transmission. Intuitively, when a sufficiently high data rate is required, all the APs should be activated to contribute to data transmission. By comparing Figs. \ref{FIG_Snapshot_Rmin3dot5}--\ref{FIG_Snapshot_Rmin0}, it can be found that as the per-user rate constraint $R_{th}$ decreases, more and more APs that are relatively far away from users become isolated, which could be switched into sleep mode to save energy. This corroborates that our clustered cell-free networking problem has successfully incorporated the sleep mode operation of APs, which is especially important for future dense wireless networks with a larger number of APs. In the case without rate performance guarantee, i.e., $R_{th}=0$, each user is simply associated to its closest AP only, as shown in Fig. \ref{FIG_Snapshot_Rmin0}, i.e., the current cellular network structure. 

\begin{figure*}[t]
	\centering
	{	\subfloat[]{\includegraphics[height=0.35\textwidth]{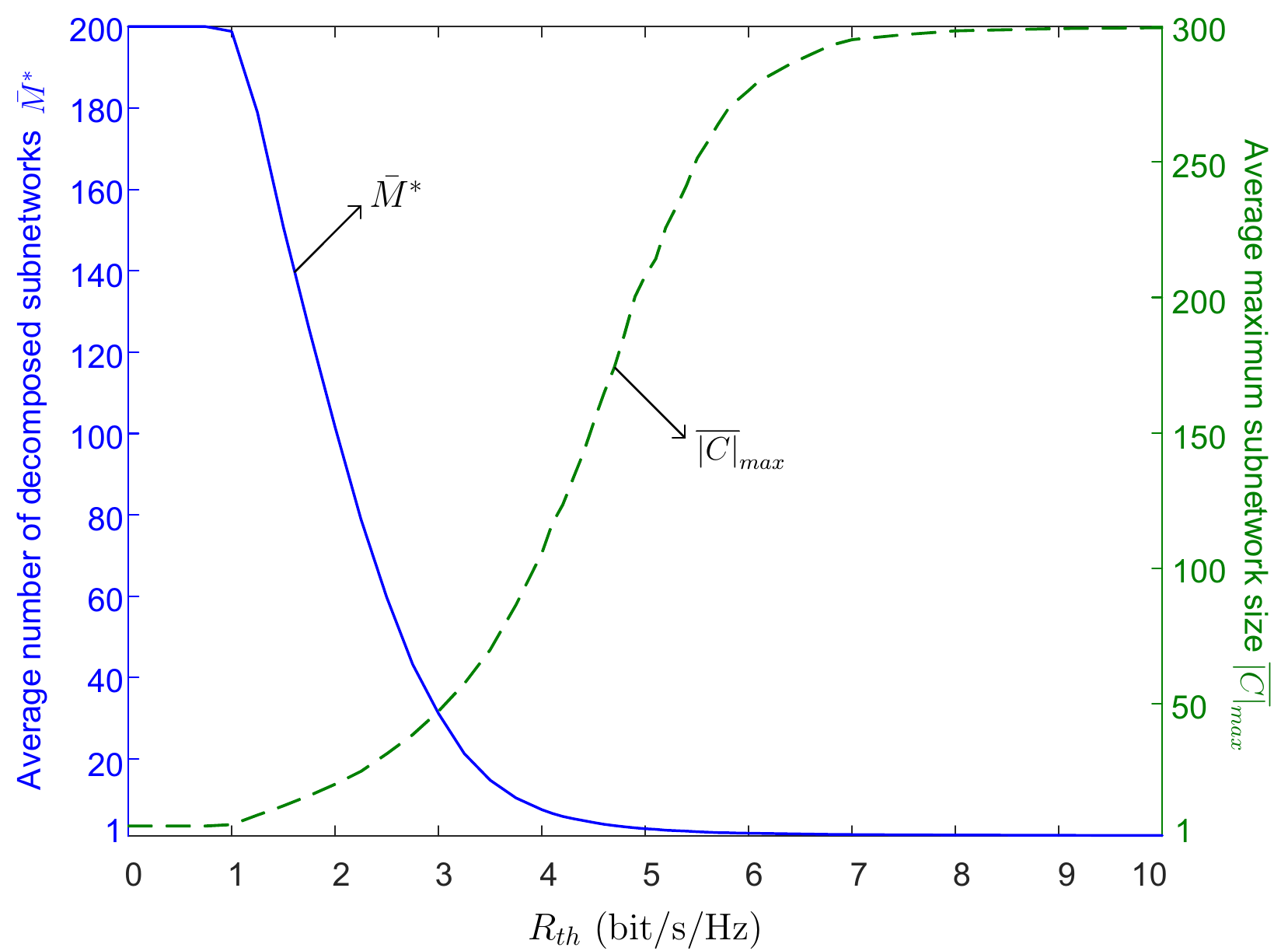}
			\label{FIG_AveM_DAS}}
		\subfloat[]{\includegraphics[height=0.35\textwidth]{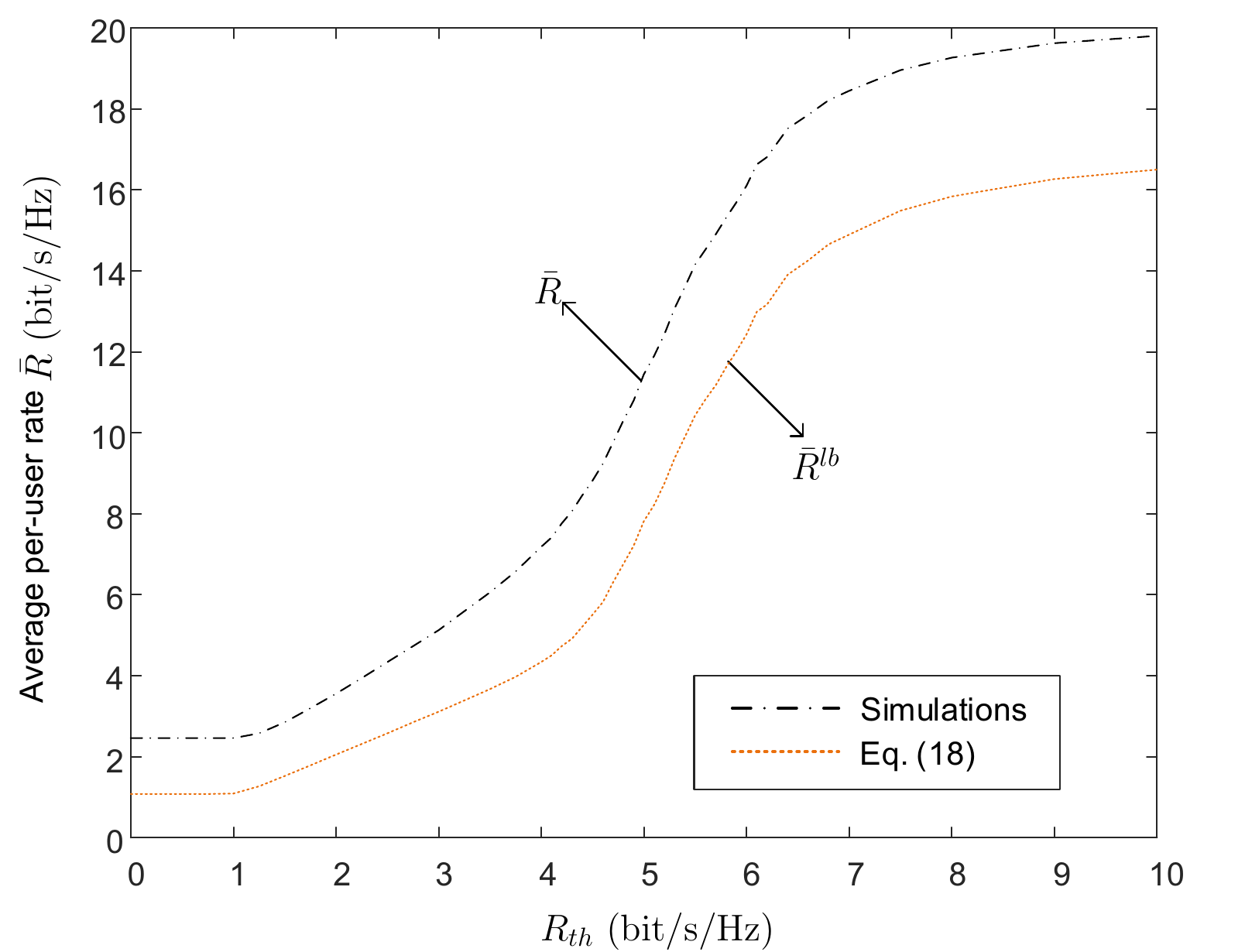}
			\label{FIG_AveR_DAS}}
	}
	\caption{(a) Average number of decomposed subnetworks $\bar{M^{*}}$ and average maximum subnetwork size $\overline{|C|}_{max}$, and (b) average per-user rate $\bar{R}$ versus the per-user rate constraint $R_{th}$ with the proposed RC-NetDecomp algorithm. $L=200$, $K=100$, $P_{t}/\sigma^2=0$dB, $\alpha=4$.}
	\label{FIG_AveMandAveR_DAS}
\end{figure*}

To have a closer look at the transition of the network structure by increasing the per-user rate constraint $R_{th}$, Fig. \ref{FIG_AveM_DAS} presents the average number of decomposed subnetworks $\bar{M^{*}}$ and the average maximum subnetwork size $\overline{|C|}_{max}$ with the RC-NetDecomp algorithm. It is shown in Fig. \ref{FIG_AveM_DAS} that as the per-user rate requirement $R_{th}$ increases from 0 to 10 bit/s/Hz, the average number of decomposed subnetworks $\bar{M^{*}}$ gradually decreases from 200 to 1. Specifically, when $R_{th}$ is smaller than the lower-bound of the per-user rate achieved under the cellular structure, $M^{*}$ remains at the number of APs $L=200$. As $R_{th}$ increases beyond that, joint processing of APs would be required for interference cancellation to improve the per-user rate performance. As a result, the number of decomposed subnetworks $M^{*}$ decreases with $R_{th}$ until $M^{*}=1$, i.e., all the APs jointly serve all the users in the network. For the same reason, the average maximum subnetwork size $\overline{|C|}_{max}$ increases with the threshold $R_{th}$, indicating increased joint processing complexity and signaling overhead. It can be clearly seen that with the proposed RC-NetDecomp algorithm, the network structure varies smoothly from the current cellular network to the fully cooperative network by increasing the per-user rate constraint $R_{th}$. 

Fig. \ref{FIG_AveR_DAS} further shows the corresponding average per-user rate $\bar{R}$ achieved and the average of the per-user rate lower-bound given in (\ref{AveR-lb}), $\bar{R}^{lb}$. Both of them are well above the per-user rate constraint $R_{th}$, indicating that the proposed RC-NetDecomp algorithm can always meet the per-user rate requirement. Since the signaling overhead and  joint processing complexity inside each subnetwork increases with the subnetwork size, it can be concluded from Fig. \ref{FIG_AveMandAveR_DAS} that the required per-user rate $R_{th}$ determines a crucial rate-complexity/signaling overhead tradeoff. 

\subsubsection{Comparison with User-Centric Clustering}

\begin{figure*}[t]
	\centering
	{	\subfloat[$S=1$]{\includegraphics[width=0.33\textwidth]{Snapshot-DAS-0.pdf}
			\label{FIG_Snapshot_DAS_V1}}
		\subfloat[$S=2$]{\includegraphics[width=0.33\textwidth]{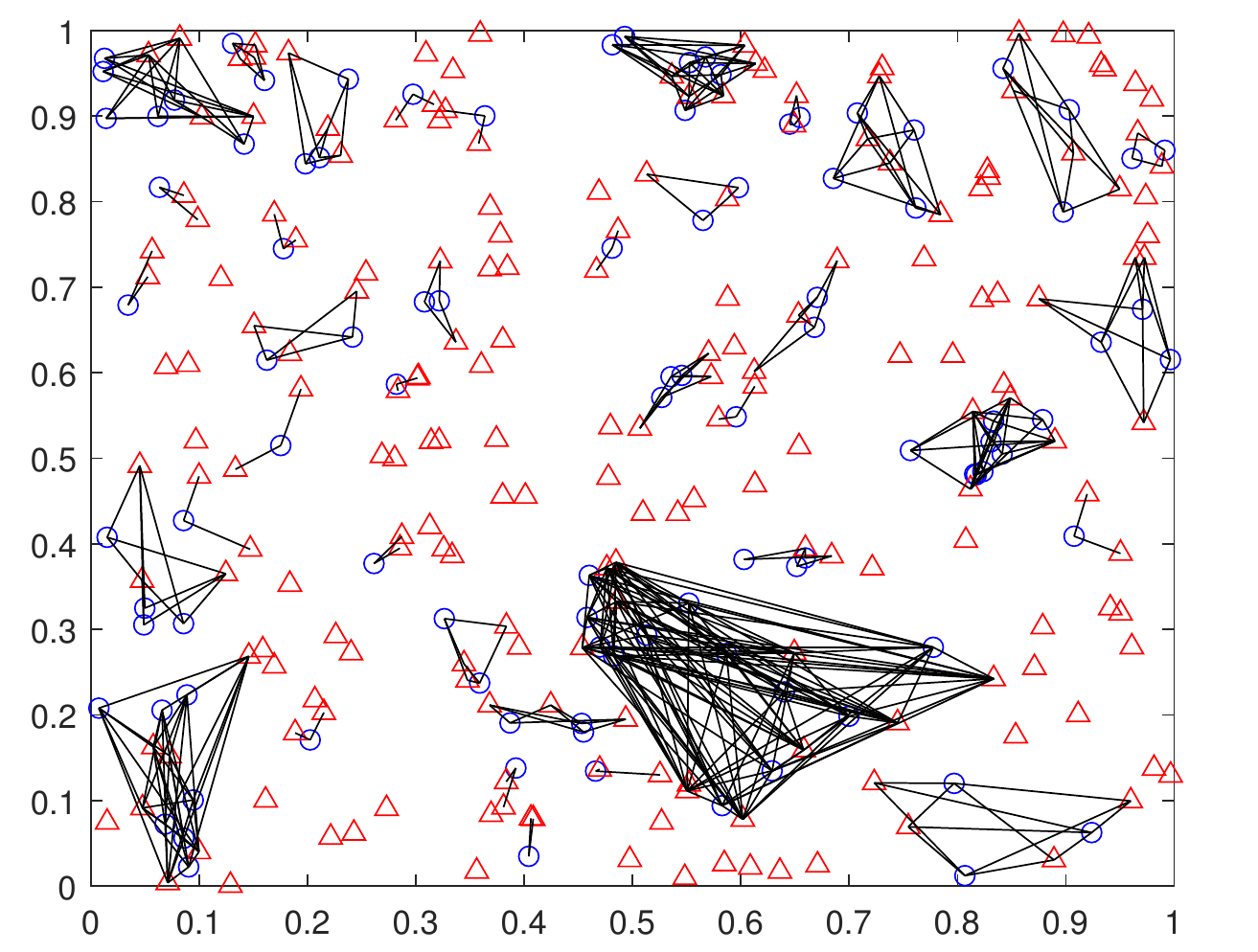}
			\label{FIG_Snapshot_DAS_V2}}
		\subfloat[$S=3$]{\includegraphics[width=0.33\textwidth]{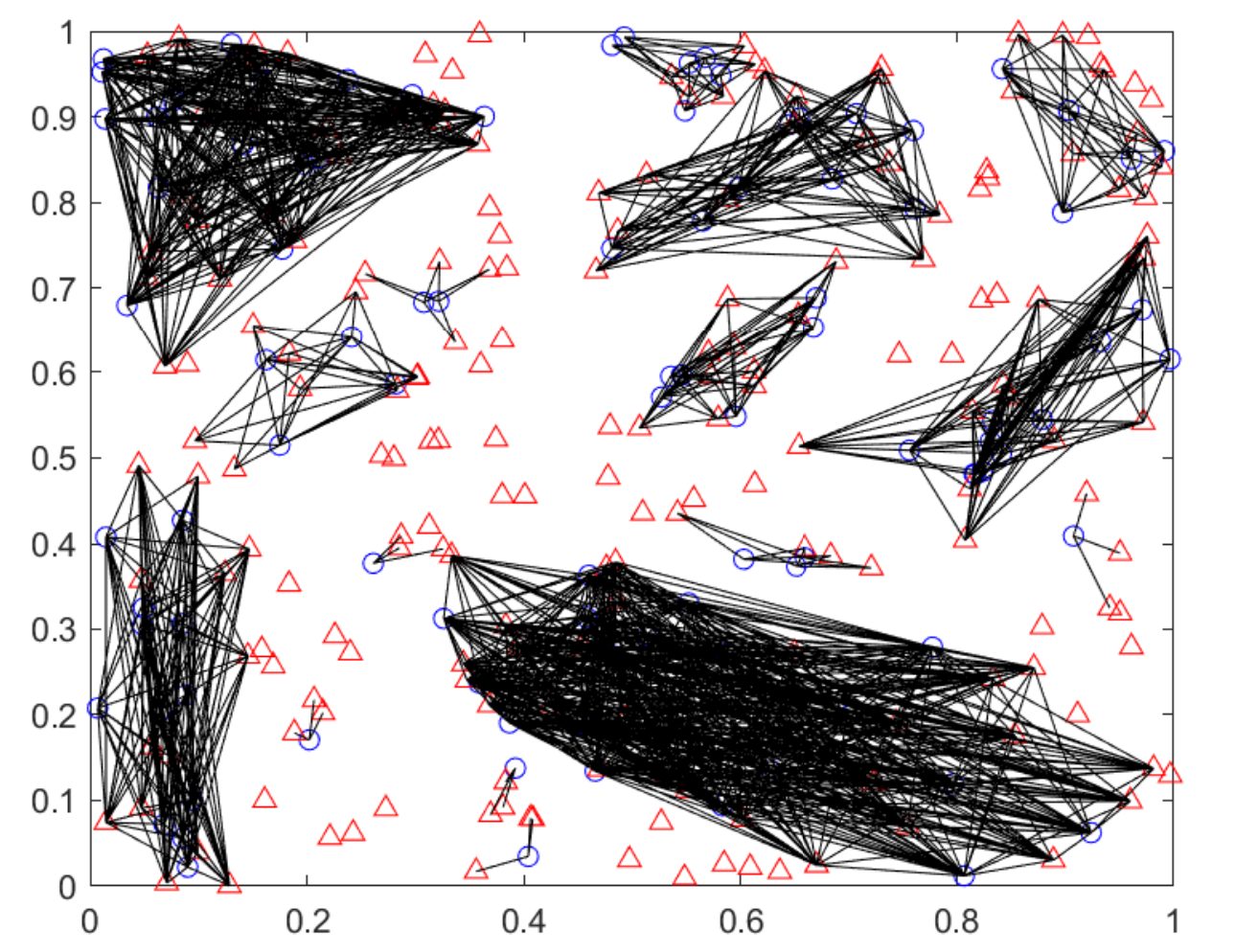}
			\label{FIG_Snapshot_DAS_V3}}
	}
	\caption{Decomposed subnetworks of the randomly generated cell-free massive MIMO system shown in Fig. \ref{FIG_Snapshot_DAS} by adopting the user-centric clustering algorithm in \cite{Junyuan_VC}. ``$\textcolor{blue}{\circ}$'' represents a user and ``$\textcolor{red}{\triangle}$'' represents an AP. $L=200$, $K=100$, $\alpha=4$.}
	\label{FIG_Snapshot_DAS_V}
\end{figure*}

With the user-centric clustering algorithm in \cite{Junyuan_VC}, Fig. \ref{FIG_Snapshot_DAS_V} presents the network decomposition results under the same topology presented in Fig. \ref{FIG_Snapshot_DAS} by varying the number of APs chosen by each user, $S$. It can be clearly seen that with $S=1$, the network structure is the same as that in Fig. \ref{FIG_Snapshot_Rmin0}, i.e., the cellular network. As the number of APs chosen by each user, $S$, increases, more and more users are grouped together for joint processing. However, a giant subnetwork appears when $S=3$, implying that the system complexity and signaling overhead could be still high after network decomposition. 

\begin{figure}[t]
	\begin{center}
		\includegraphics[width=0.6\textwidth]{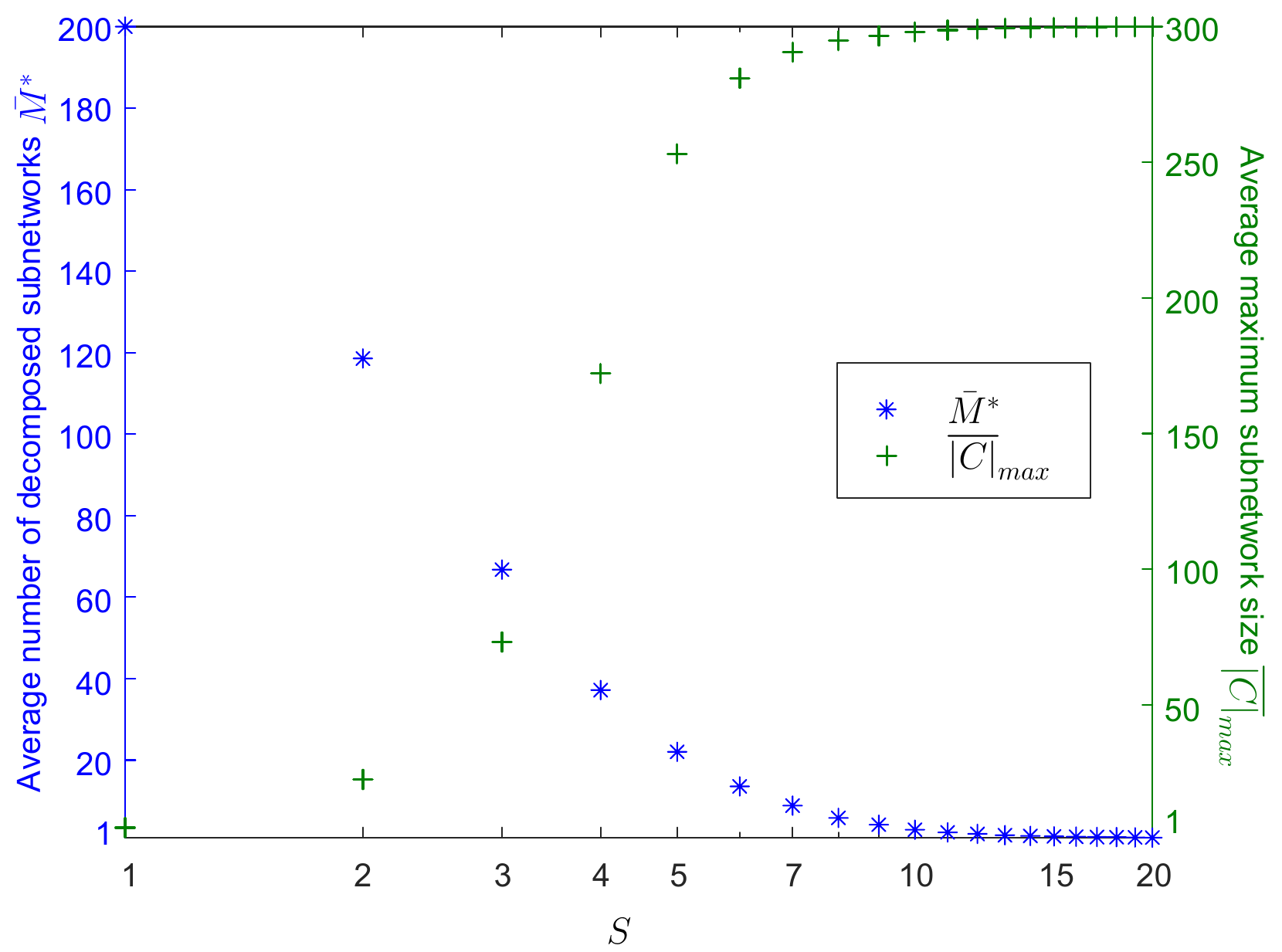}
		\caption{Average number of decomposed subnetworks $\bar{M}^{*}$ and average maximum subnetwork size $\overline{|C|}_{max}$ versus the number of APs chosen by each user, $S$, with the user-centric clustering algorithm in \cite{Junyuan_VC}. $L=200$, $K=100$, $\alpha=4$.}
		\label{FIG_AveMAveS_DAS_V}
	\end{center}
\end{figure}

Fig. \ref{FIG_AveMAveS_DAS_V} further presents the average number of decomposed subnetworks\footnote{To align with the assumptions in this paper, an isolated AP is counted as a single subnetwork.} $\bar{M^{*}}$ and the average maximum subnetwork size $\overline{|C|}_{max}$ by increasing the number of APs chosen by each user, $S$, from 1 to $200$. A similar network structure transition can be observed by tuning $S$. However, only a small number of options are available as marked in Fig. \ref{FIG_AveMAveS_DAS_V}, since $S$ needs to be an integer. Moreover, the average number of decomposed subnetworks $\bar{M^{*}}$ drops quickly as $S$ increases. By comparing Fig. \ref{FIG_AveMAveS_DAS_V} with Fig. \ref{FIG_AveM_DAS}, we can see that with the same $\bar{M^{*}}$, the average maximum subnetwork size  $\overline{|C|}_{max}$ with the user-centric clustering algorithm in \cite{Junyuan_VC} is much larger than that with the RC-NetDecomp algorithm. For instance, with $\bar{M^{*}}=67$ (i.e., $S$ = 3 in
Fig. \ref{FIG_AveMAveS_DAS_V}), $\overline{|C|}_{max}=73$ with the user-centric clustering algorithm. By contrast, with the RC-NetDecomp algorithm, the average maximum subnetwork size $\overline{|C|}_{max}$ is significantly reduced to 30, indicating much lower complexity and signaling overhead for joint processing. It can be concluded that compared to the user-centric clustering benchmark, the proposed RC-NetDecomp algorithm enables a much finer-grained tuning of subnetwork size and more efficient reduction of system complexity and signaling overhead.  

\begin{figure*}[t]
	\centering
	{	\subfloat[$M=9$]{\includegraphics[width=0.33\textwidth]{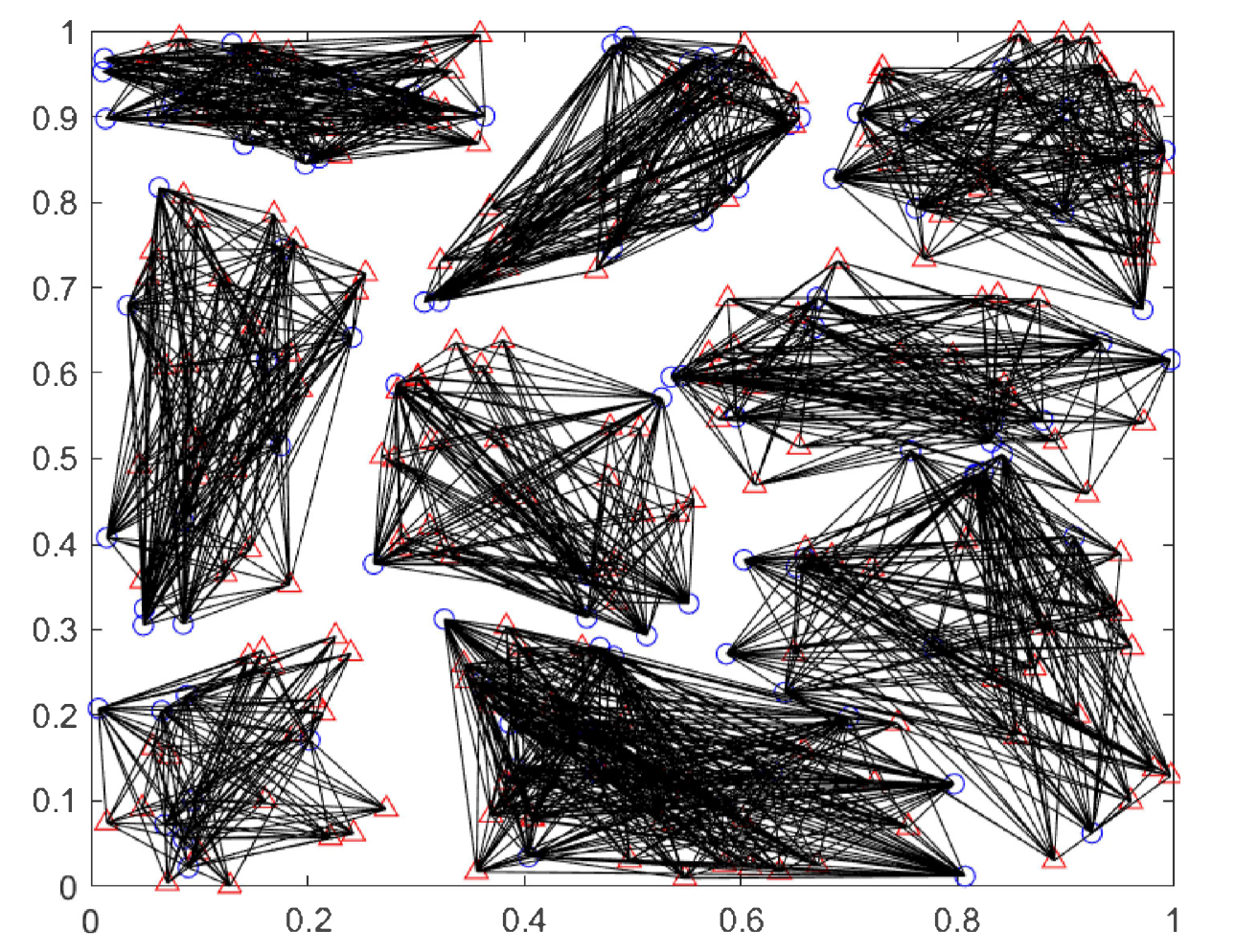}
			\label{FIG_Snapshot_DAS_Gold_M9}}
		\subfloat[$M=88$]{\includegraphics[width=0.33\textwidth]{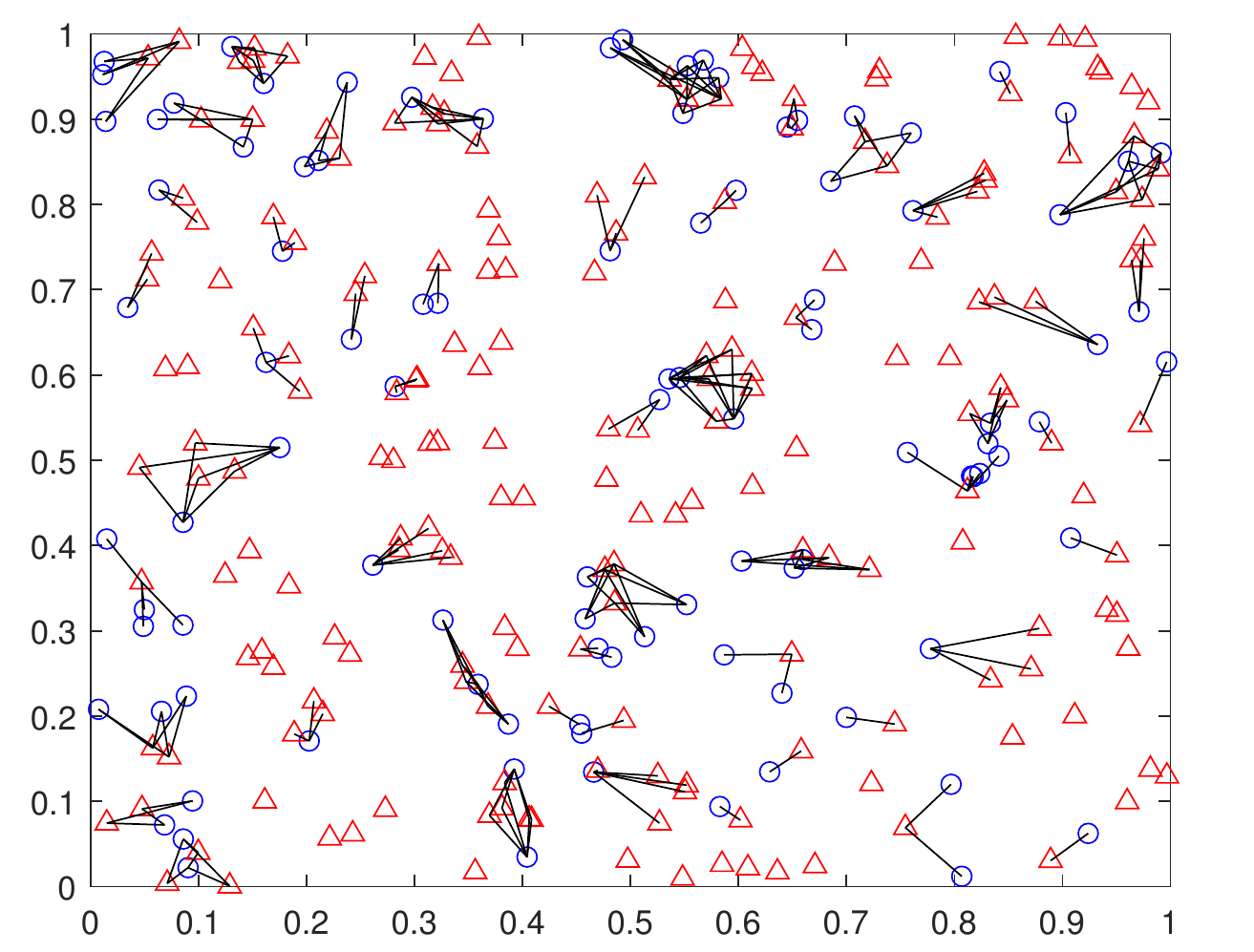}
			\label{FIG_Snapshot_DAS_Gold_M88}}
		\subfloat[$M=200$]{\includegraphics[width=0.33\textwidth]{Snapshot-DAS-0.pdf}
			\label{FIG_Snapshot_DAS_Gold_M200}}
	}
	\caption{Decomposed subnetworks of the randomly generated cell-free massive MIMO system shown in Fig. \ref{FIG_Snapshot_DAS} by adopting the AP-centric clustering algorithm in \cite{Goldsmith}. ``$\textcolor{blue}{\circ}$'' represents a user and ``$\textcolor{red}{\triangle}$'' represents an AP. $L=200$, $K=100$, $\alpha=4$.}
	\label{FIG_Snapshot_DAS_Gold}
\end{figure*}

\subsubsection{Comparison with AP-Centric Clustering}
With the AP-centric clustering algorithm proposed in \cite{Goldsmith}, Fig. \ref{FIG_Snapshot_DAS_Gold} presents the snapshots of the decomposed subnetworks under the same topology shown in Fig. \ref{FIG_Snapshot_DAS} with various numbers of decomposed subnetworks $M$. It can be clearly seen from Fig. \ref{FIG_Snapshot_DAS_Gold} that the clustered cell-free network structures with the AP-centric clustering algorithm in \cite{Goldsmith} are similar to those with the proposed RC-NetDecomp algorithm, indicating that the AP-centric approach can also enable flexible network decomposition. However, there could be a large number of users located at the subnetwork edge, who would suffer from strong inter-subnetwork interference. By contrast, with the proposed RC-NetDecomp algorithm, it is the APs that are usually located at the subnetwork edge as the edge weights on graph $G$ are modeled in a user-centric way. As a result, higher per-user rate could be expected with the proposed RC-NetDecomp algorithm.

\begin{figure*}[t]
	\centering
	{	\subfloat[]{\includegraphics[width=0.45\textwidth]{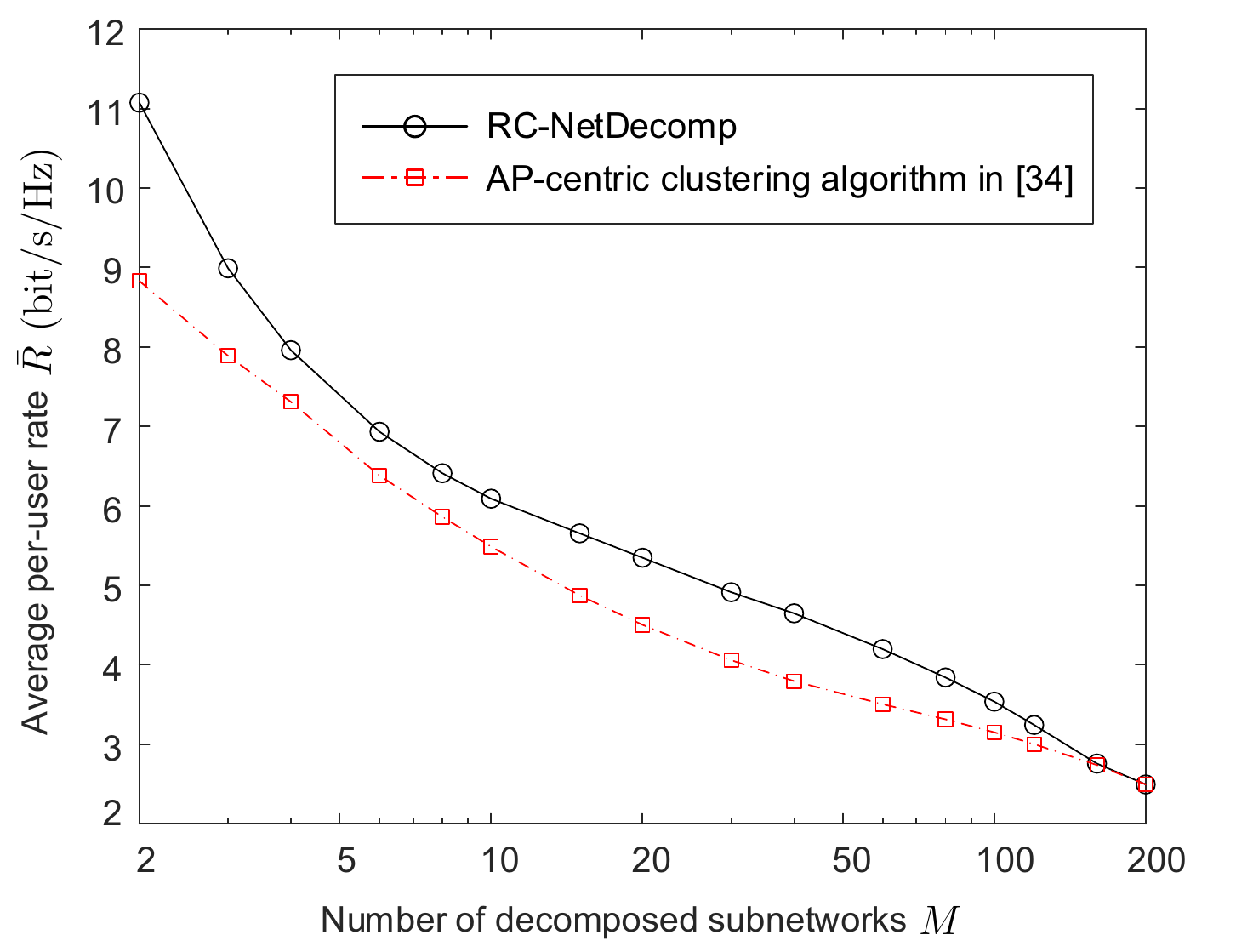}
			\label{FIG_AveR_DAS_all}}
		\subfloat[]{\includegraphics[width=0.45\textwidth]{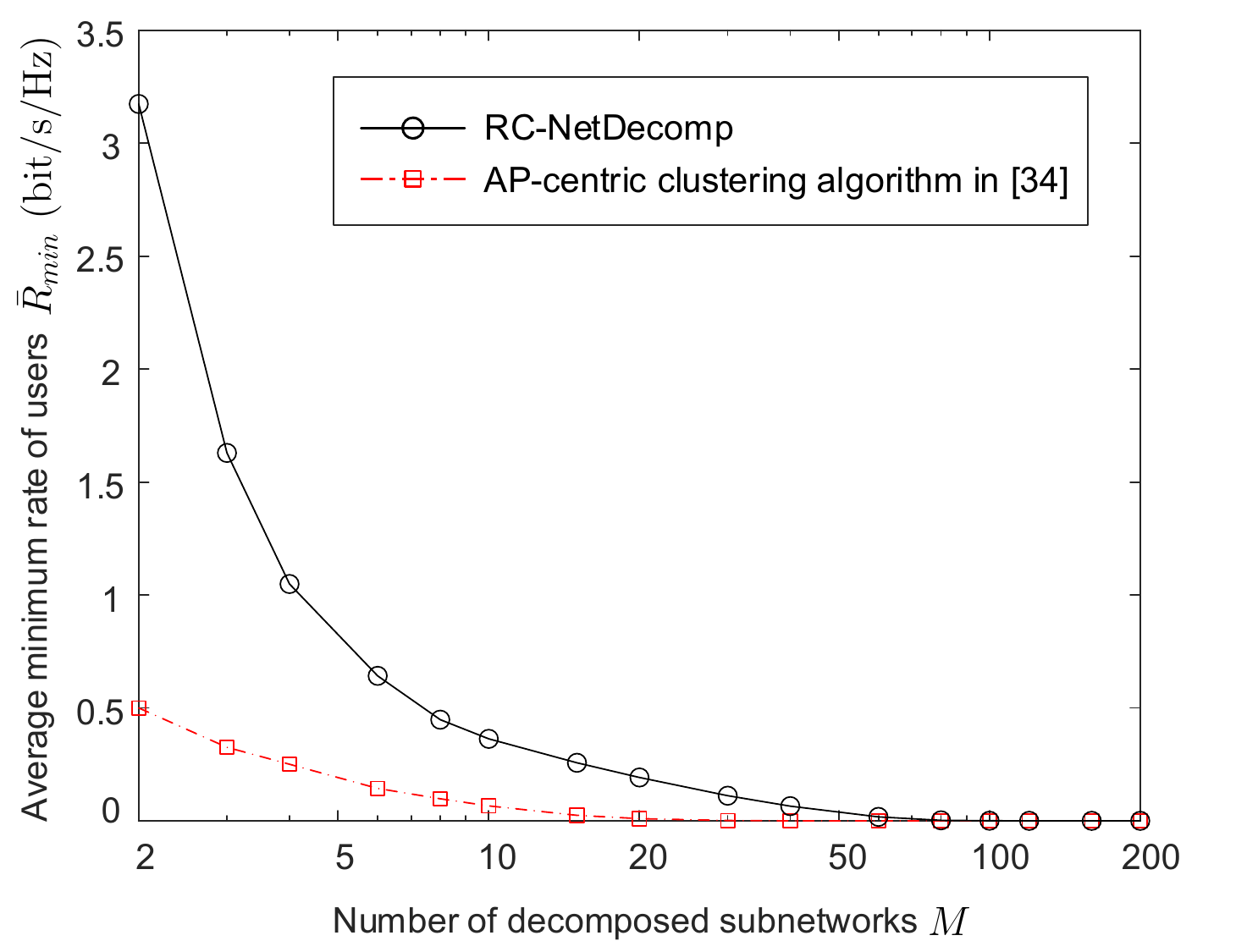}
			\label{FIG_AveRmin_DAS_all}} \\
		\subfloat[]{\includegraphics[width=0.45\textwidth]{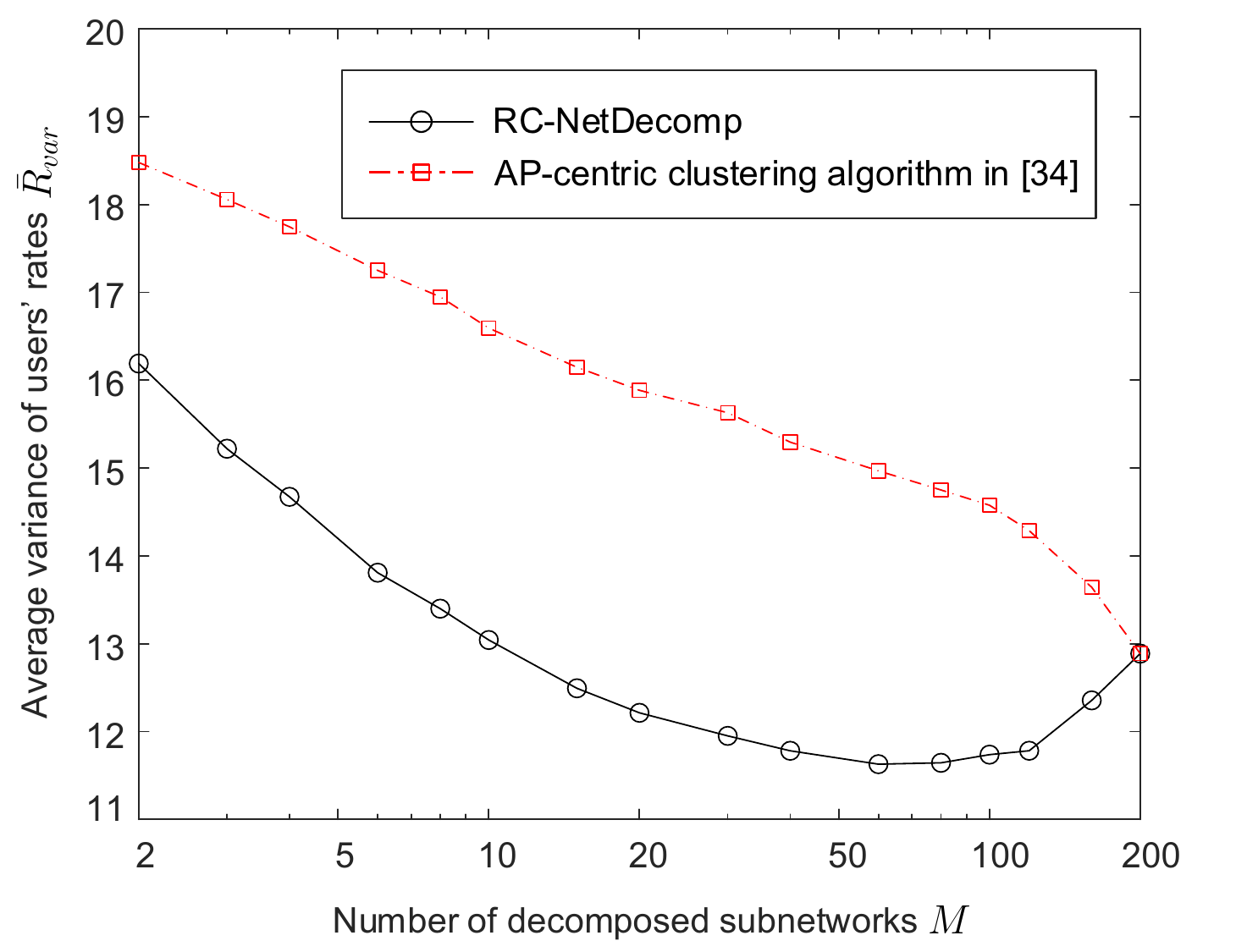}
			\label{FIG_AveRvar_DAS_all}}
		\subfloat[]{\includegraphics[width=0.45\textwidth]{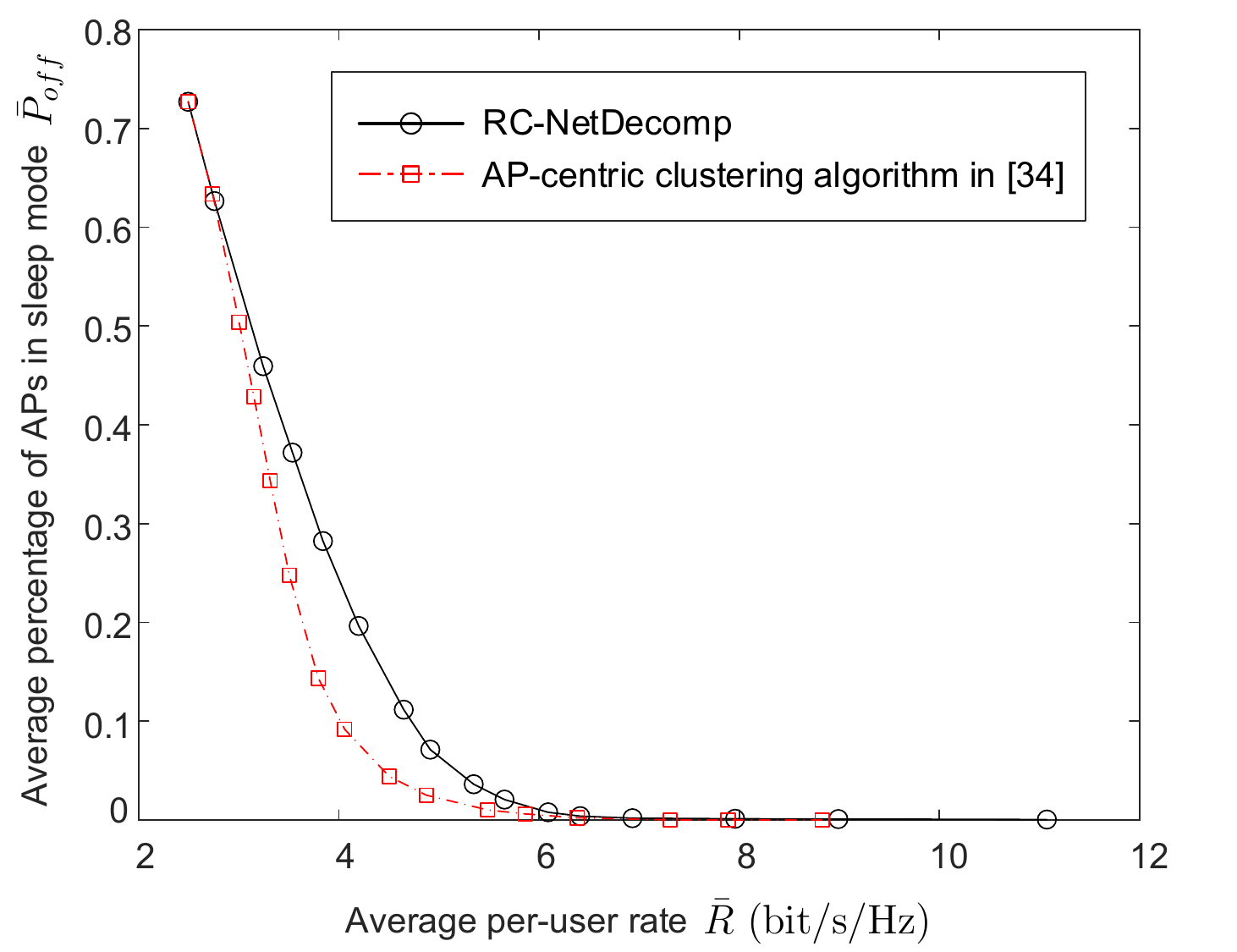}
			\label{FIG_AveIsoAp_DAS_all}}	
	}
	\caption{(a) Average per-user rate $\bar{R}$, (b) average minimum rate of users $\bar{R}_{min}$, (c) average variance of users' rates $\bar{R}_{var}$ and (d) average percentage of APs in sleep mode $\bar{P}_{off}$ with the proposed RC-NetDecomp algorithm and the AP-centric clustering algorithm in \cite{Goldsmith}. $L=200$, $K=100$, $P_{t}/\sigma^2=0$dB, $\alpha=4$.}
	\label{FIG_Ave_DAS}
\end{figure*}

Figs. \ref{FIG_AveR_DAS_all}--\ref{FIG_AveRvar_DAS_all} present the average per-user rate $\bar{R}$, the average minimum rate of users $\bar{R}_{min}$ and the average variance of users' rates $\bar{R}_{var}$  with the proposed RC-NetDecomp algorithm and the AP-centric clustering algorithm in \cite{Goldsmith}. For the proposed RC-NetDecomp algorithm, it can be observed that both the average per-user rate $\bar{R}$ and the average minimum rate $\bar{R}_{min}$ decrease as the number of subnetworks $M$ increases, yet the average variance of users' rates $\bar{R}_{min}$ first decreases and then increases when $M$ exceeds 80. To understand why the average rate variance increases with $M$ when $M$ is large, note from Fig. \ref{FIG_AveRmin_DAS_all} that the average minimum rate $\bar{R}_{min}$ drops to zero at $M=80$. With $M$ further increasing, more and more users are in outage, thus leading to higher average variance of users' rates, as observed in Fig. \ref{FIG_AveRvar_DAS_all}.

For the comparision with the AP-centric algorithm in \cite{Goldsmith}, it can be seen from Figs. \ref{FIG_AveR_DAS_all}--\ref{FIG_AveRvar_DAS_all} that the proposed RC-NetDecomp algorithm achieves higher average per-user rate $\bar{R}$, higher average minimum rate of users $\bar{R}_{min}$ and smaller average variance of users' rates $\bar{R}_{var}$, where the latter two indicate that more uniform rate performance could be expected by adopting the proposed RC-NetDecomp algorithm. It can be concluded that compared to the AP-centric clustering benchmark, our proposed RC-NetDecomp algorithm improves not only per-user rate performance but also user fairness. The comparison highlights the importance of clustering based on the locations of both APs and users. 

In addition, Fig. \ref{FIG_AveIsoAp_DAS_all} presents the average percentage of APs in sleep mode $\bar{P}_{off}$ with both RC-NetDecomp and the AP-centric clustering algorithm in \cite{Goldsmith}. We can see that given the same average per-user rate $\bar{R}$, the average percentage of APs in sleep mode $\bar{P}_{off}$ with the proposed RC-NetDecomp algorithm is much higher than that with the AP-centric clustering algorithm in \cite{Goldsmith} when the average per-user rate $\bar{R}$ is between 3 bit/s/Hz and 6 bit/s/Hz, which indicates significant improvement in energy efficiency by adopting our RC-NetDecomp algorithm.  

\subsection{Special Case of Single-Cell Massive MIMO with $L=1$ }
With the number of APs $L=1$, all the $N$ beams are available at the same AP and the general cell-free wireless network considered in this paper reduces to a fixed-beam based single-cell massive MIMO system. In such a special case, the distance $d_{k,n}$ is independent of the beam index $b_{n}$. As a result, the edge weight given in (\ref{w}) reduces to $w_{k,n}=D_{k,n}/D_{k,n_{k}^{*}}$,
%\begin{align}\label{w-beam}
%	w_{k,n}=\frac{D_{k,n}}{D_{k,n_{k}^{*}}},
%\end{align}
where $D_{k,n_{k}^{*}}$ is equal to the largest beam gain of user $u_{k}$, i.e., $D_{k,n_{k}^{*}}=\max_{n=1, 2, \cdots N} D_{k,n}$. For simplicity, let us assume that the AP is located at the center of a circular cell and a linear antenna array of $N$ antenna elements spaced at half wavelength is placed horizontally at the AP with discrete Fourier transform (DFT) beamforming adopted to form $N$ fixed beams. Without loss of generality, by assuming a line-of-sight (LoS) channel at mmWave frequencies, the beam gain of beam $n$ numbered from the left-hand side to the right-hand side naturally for user $u_{k}$ is given by $D_{k,n}=\frac{\sin^2\left(\frac{N\pi}{2} \cos \theta_{k}-\psi_{n}\right)}{N\sin^2\left(\frac{\pi}{2}  \cos \theta_{k}-\frac{1}{N}\psi_{n}\right)}$, where $\theta_{k}$ denotes the angle of departure (AoD) of the signal received at user $u_{k}$, and $\psi_{n}=\left(-\frac{N+1}{2}+n\right)\pi$ \cite{Junyuan_LBA}.

\begin{figure*}[t]
	\centering
	{\subfloat[$R_{th}=5$ bit/s/Hz]{\includegraphics[width=0.24\textwidth]{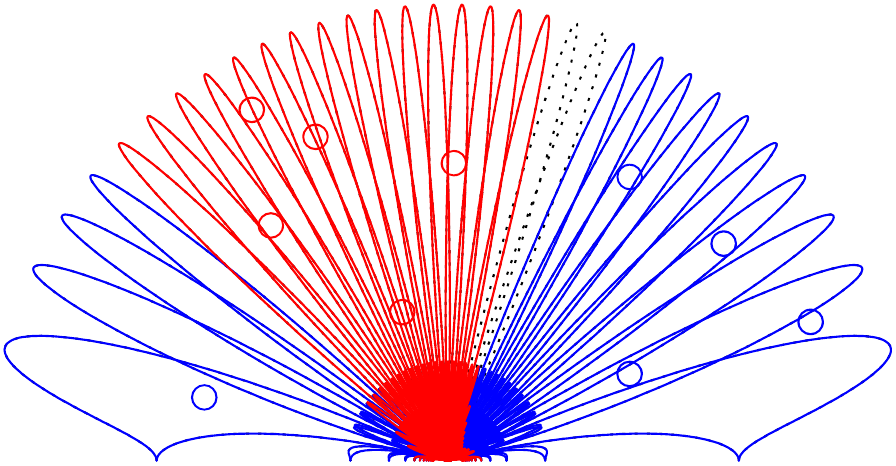}
			\label{FIG_Snapshot_Beam_Rmin5}}
		\subfloat[$R_{th}=4$ bit/s/Hz]{\includegraphics[width=0.24\textwidth]{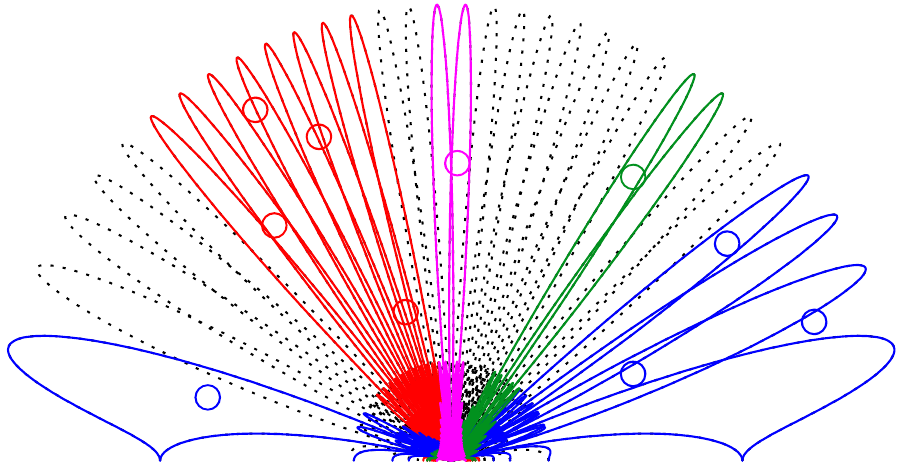}
			\label{FIG_Snapshot_Beam_Rmin4}} 
		\subfloat[$R_{th}=3$ bit/s/Hz]{\includegraphics[width=0.24\textwidth]{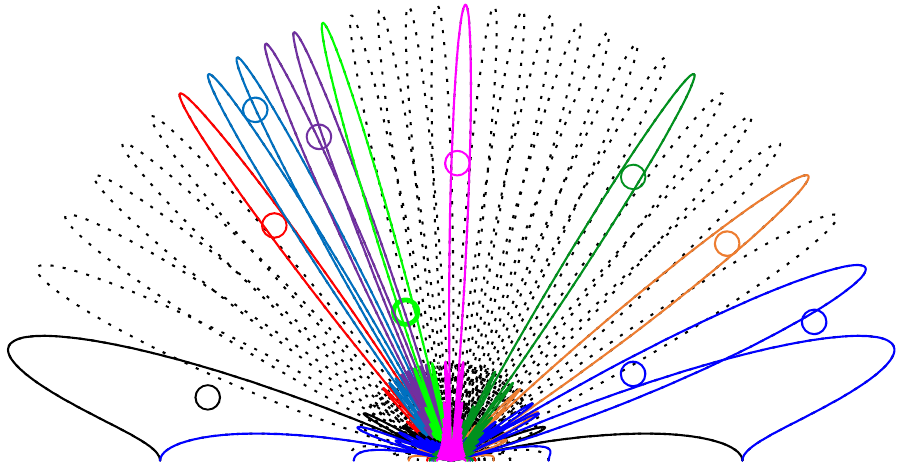}
			\label{FIG_Snapshot_Beam_Rmin3}}
		\subfloat[$R_{th}=0$]{\includegraphics[width=0.24\textwidth]{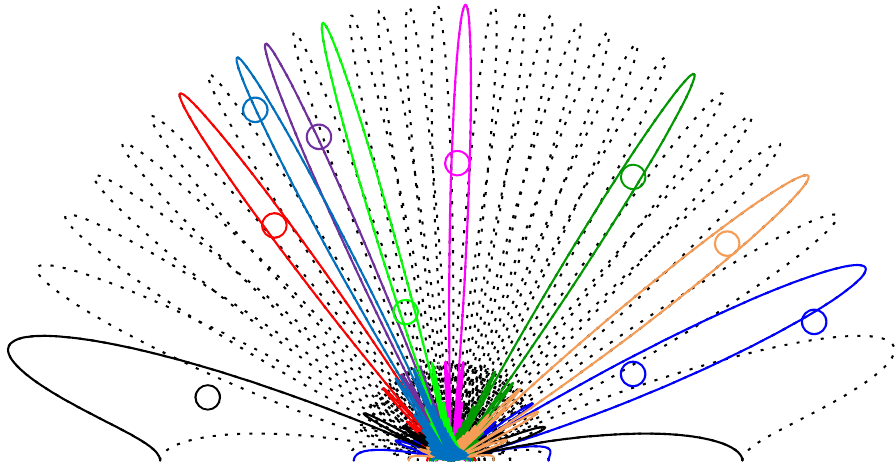}
			\label{FIG_Snapshot_Beam_Rmin0}}
	}
	\caption{Decomposition results of a fixed-beam based single-cell massive MIMO system by adopting the proposed RC-NetDecomp algorithm. ``$\circ$'' represents a user. Users and beams in the same color are in the same subnetwork. Inactive beams are drawn in dashed lines. $N=32$, $K=10$.}
	\label{FIG_Snapshot_Beam}
\end{figure*} 

\subsubsection{Decomposition Results}
Fig. \ref{FIG_Snapshot_Beam} illustrates the decomposition results of a fixed beam based single-cell massive MIMO system with the proposed RC-NetDecomp algorithm when the per-user rate constraint $R_{th}$ is $5$ bit/s/Hz, $4$ bit/s/Hz, $3$ bit/s/Hz and $0$. Similar to the cell-free massive MIMO case, as the per-user rate constraint $R_{th}$ increases, more and more beams are grouped together to serve users jointly for the sake of inter-beam interference cancellation. In the scenario with very low or even no rate requirement, a beam is associated with the users falling into its beam coverage as depicted in Fig. \ref{FIG_Snapshot_Beam_Rmin0}, reducing to the special case of beam allocation based massive MIMO system investigated in \cite{Junyuan_LBA}, where a user is served by a single beam independently. In addition, it can be seen that with the proposed RC-NetDecomp algorithm, there are beams associated with no users, which can be turned off to save not only transmit power but also the energy consumed by radio frequency (RF) chains, as the number of RF chains in use is now smaller than the total number of beams $N$.  

\begin{figure*}[t]
	\centering
	{	\subfloat[]{\includegraphics[height=0.35\textwidth]{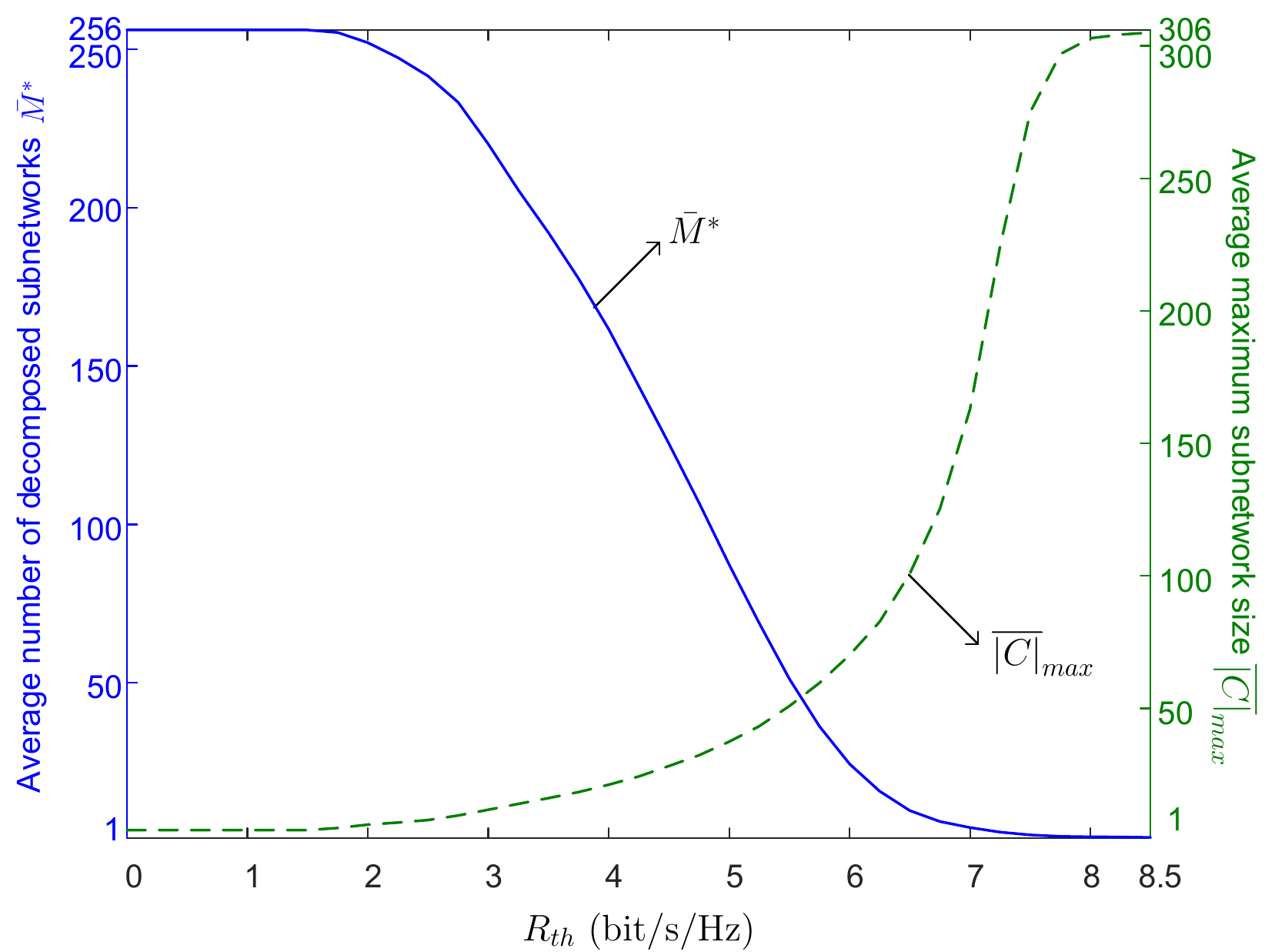}
			\label{FIG_AveMAveS_Beam}}
		\subfloat[]{\includegraphics[height=0.35\textwidth]{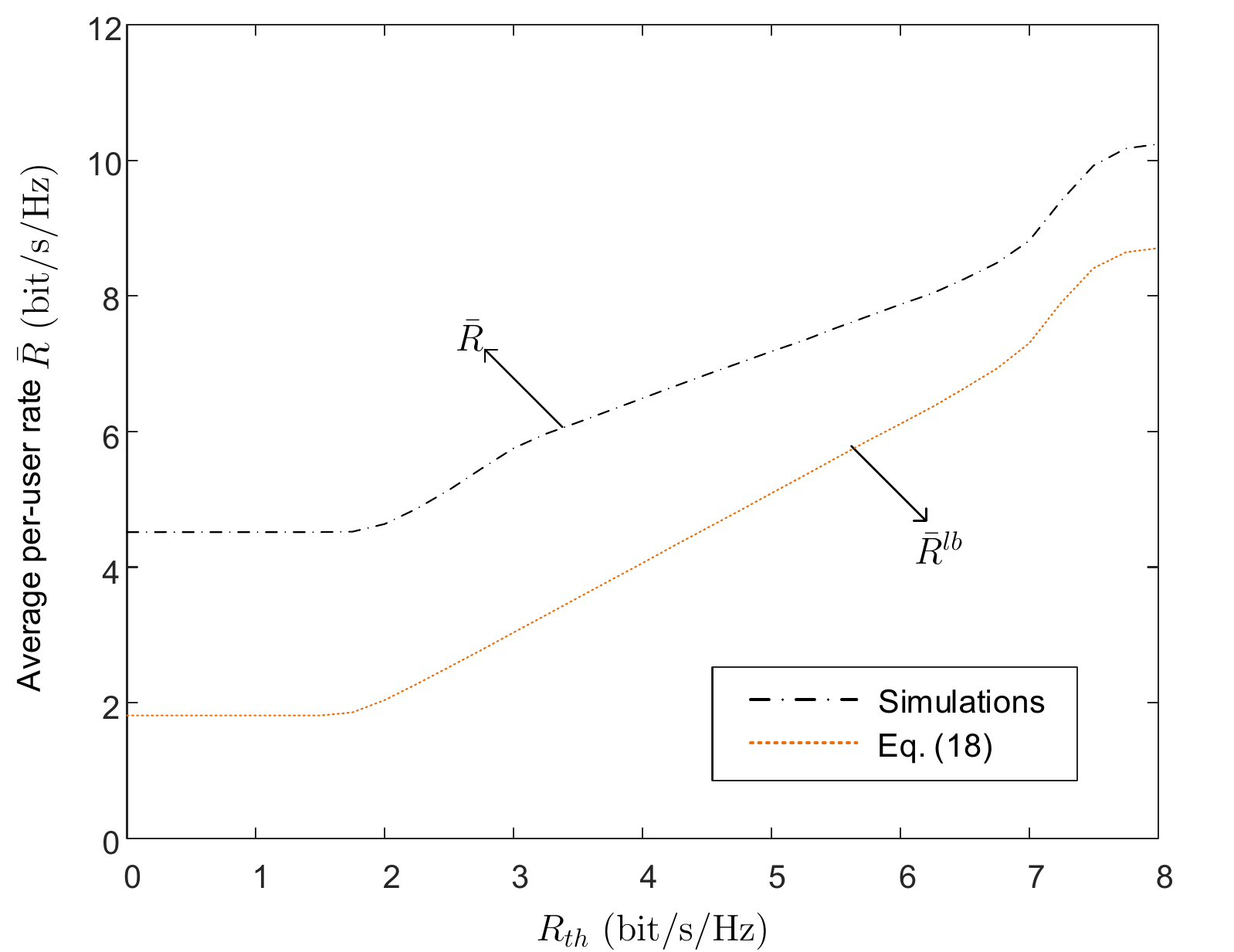}
 			\label{FIG_AveR_Beam}}
	}
	\caption{(a) Average number of decomposed subnetworks $\bar{M^{*}}$ and average maximum subnetwork size $\overline{|C|}_{max}$, and (b) average per-user rate $\bar{R}$ versus the per-user rate constraint $R_{th}$ with the proposed RC-NetDecomp algorithm. $N=256$, $K=50$, $P_{t}/\sigma^2=0$dB, $\alpha=2.7$.}
	\label{FIG_AveMandAveR_Beam}
\end{figure*}

Fig. \ref{FIG_AveMandAveR_Beam} presents the average number of subnetworks $\bar{M^{*}}$ together with the average maximum subnetwork size $\overline{|C|}_{max}$, and the average per-user rate $\bar{R}$ by adopting the proposed RC-NetDecomp algorithm. The average per-user rate lower-bound $\bar{R}^{lb}$ is also plotted in Fig. \ref{FIG_AveR_Beam} for comparison. Simulation results corroborate that the proposed RC-NetDecomp algorithm can always meet the per-user rate requirement. Similar to the special case of cell-free massive MIMO, we can see from Fig. \ref{FIG_AveMAveS_Beam} that the rate constraint $R_{th}$ could be carefully tuned to balance the per-user rate performance and the system complexity/signaling overhead in a fine-grained manner. 

\subsubsection{Comparison with User-Centric Clustering}
\begin{figure*}[t]
	\centering
	{	\subfloat[$S=1$]{\includegraphics[width=0.24\textwidth]{Snapshot-Beam-0.pdf}
			\label{FIG_Snapshot_Beam_V1}}
		\subfloat[$S=2$]{\includegraphics[width=0.24\textwidth]{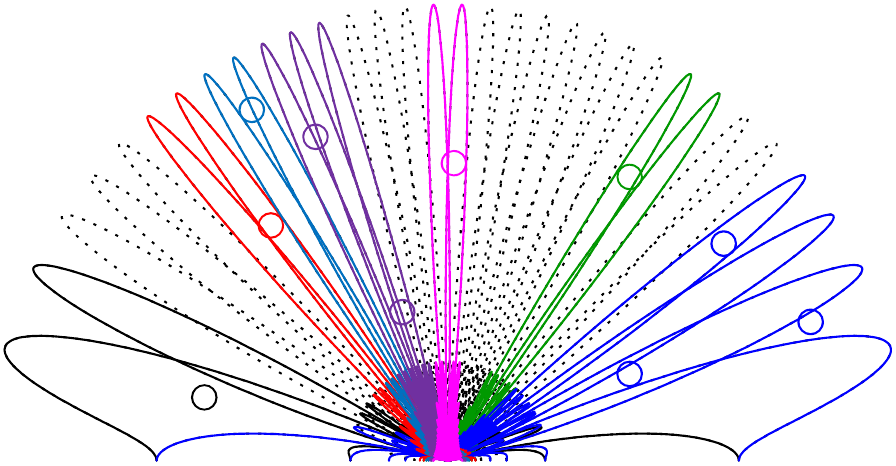}
			\label{FIG_Snapshot_Beam_V2}}
		\subfloat[$S=3$]{\includegraphics[width=0.24\textwidth]{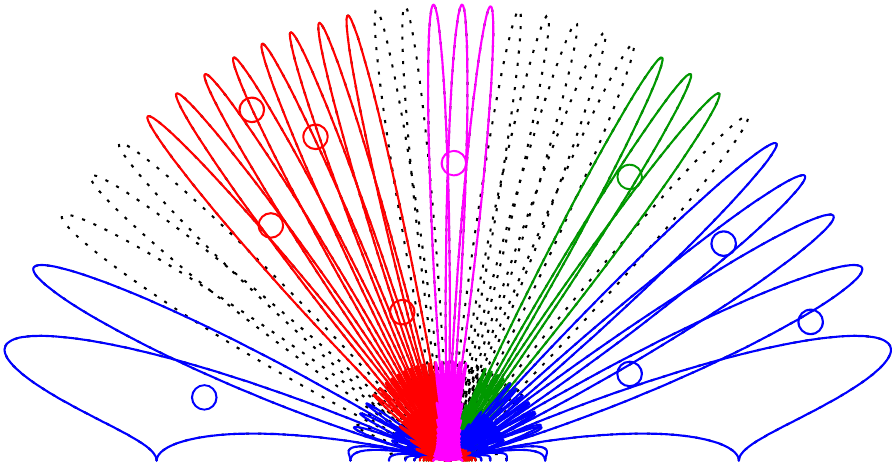}
			\label{FIG_Snapshot_Beam_V3}}
		\subfloat[$S=5$]{\includegraphics[width=0.24\textwidth]{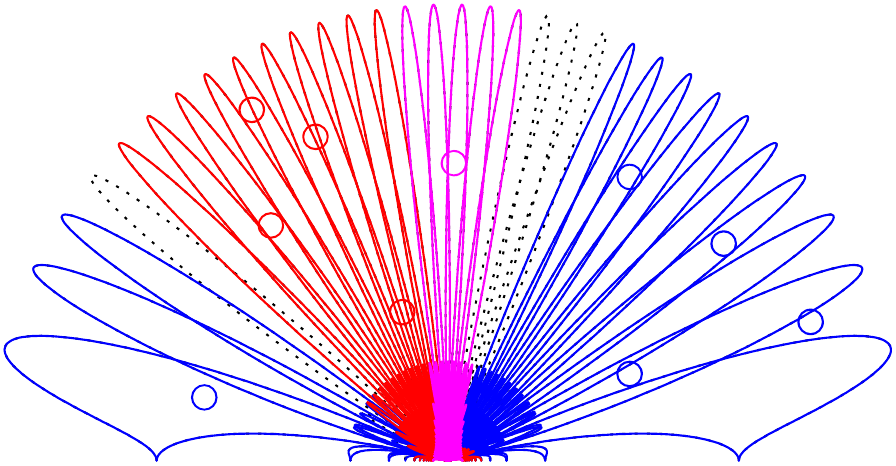}
			\label{FIG_Snapshot_Beam_V5}} 
	}
	\caption{Decomposition results of a fixed-beam based single-cell massive MIMO system with the user-centric clustering algorithm in \cite{Junyuan_VC}. ``$\circ$'' represents a user. Users and beams in the same color belong to the same subnetwork. Inactive beams are drawn in dashed lines. $N=32$, $K=10$.}
	\label{FIG_Snapshot_Beam_V}
\end{figure*}

By applying the user-centric clustering algorithm in \cite{Junyuan_VC} in the single-cell massive MIMO system, each user first selects its $S$ best beams with the highest beam gains and then group the users with overlapped associated beams, which is the same as the first step of the beam-user grouping algorithm proposed in \cite{SheuJS}. Fig. \ref{FIG_Snapshot_Beam_V} presents the decomposition results with the user-centric clustering algorithm in \cite{Junyuan_VC} under the same user locations shown in Fig. \ref{FIG_Snapshot_Beam}. By comparing Fig. \ref{FIG_Snapshot_Beam} and Fig. \ref{FIG_Snapshot_Beam_V}, it can be found that the proposed RC-NetDecomp algorithm enables much more flexible user-beam association than the user-centric clustering algorithm. For instance, in Fig. \ref{FIG_Snapshot_Beam_Rmin3}, a user close to the main direction of its best beam is served by this beam only, while a user located at the angular edge of two adjacent beams is served by these two beams jointly. By contrast, with the user-centric clustering algorithm, the number of beams chosen by each user is a constant, which might lead to unnecessary beams being activated. As illustrated in Fig. \ref{FIG_Snapshot_Beam_V2}, for example, a user close to the main direction of a beam is still served by two beams, which is unnecessary. 

\begin{figure}[t]
	\begin{center}
		\includegraphics[width=0.6\textwidth]{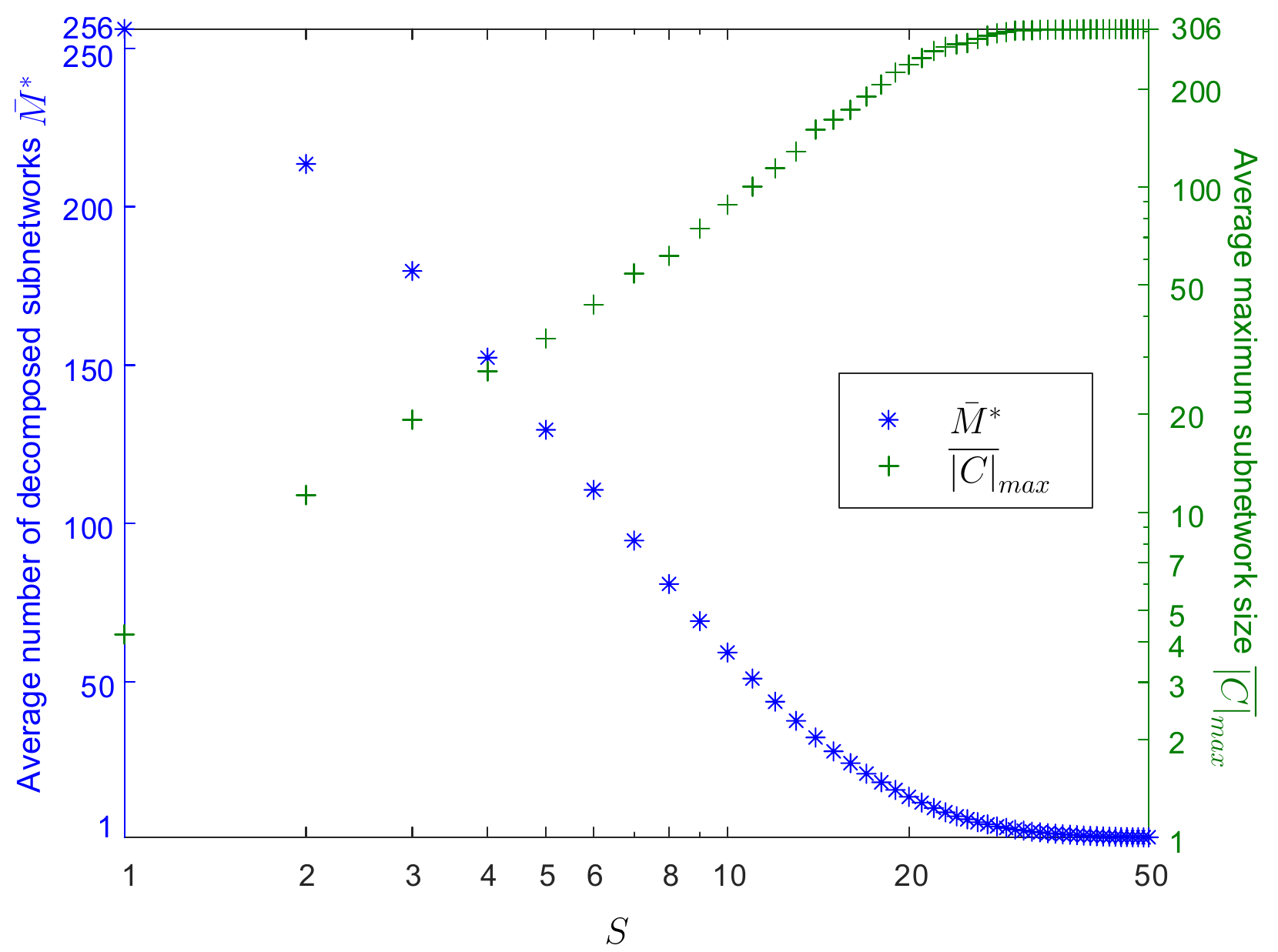}
		\caption{Average number of decomposed subnetworks $\bar{M^{*}}$ and average maximum subnetwork size $\overline{|C|}_{max}$ with the user-centric clustering algorithm in \cite{Junyuan_VC}. $N =256$, $K=50$.}
		\label{FIG_AveMAveS_Beam_V}
	\end{center}
\end{figure}

Fig. \ref{FIG_AveMAveS_Beam_V} presents the average number of decomposed subnetworks $\bar{M^{*}}$ and the average maximum subnetwork size $\overline{|C|}_{max}$ by varying the number of beams chosen by each user $S$. It can be clearly seen in this figure that the drawbacks of the user-centric approach identified in the cell-free massive MIMO special case remain here in the single-cell massive MIMO scenario. Specifically, as shown in Fig. \ref{FIG_AveMAveS_Beam_V}, the average maximum subnetwork size $\overline{|C|}_{max}$ increases dramatically with $S$, implying high joint processing complexity and signaling overhead. More importantly, there are only a small number of options to adjust the number of decomposed subnetworks and the subnetwork size, which limits the practical use of the user-centric clustering benchmark for clustered cell-free networking. 

\subsubsection{Comparison with AP-Centric Clustering}

\begin{figure*}[t]
	\centering
	{	\subfloat[$M=5$]{\includegraphics[width=0.25\textwidth]{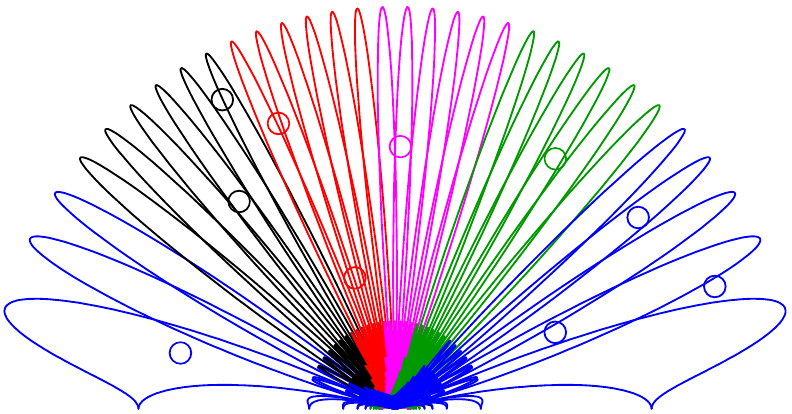}
			\label{FIG_Snapshot_DAS_Gold_M5}}
		\subfloat[$M=10$]{\includegraphics[width=0.25\textwidth]{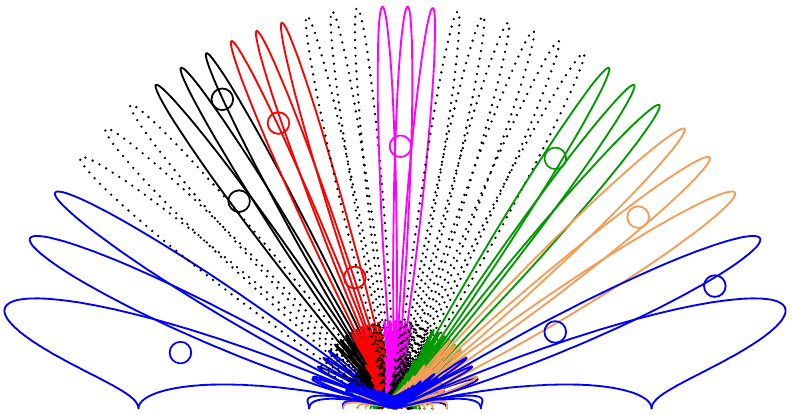}
			\label{FIG_Snapshot_DAS_Gold_M10}}
		\subfloat[$M=32$]{\includegraphics[width=0.25\textwidth]{Snapshot-Beam-0.pdf}
			\label{FIG_Snapshot_DAS_Gold_M32}}
	}
	\caption{Decomposition results of a fixed-beam based single-cell massive MIMO system with the AP-centric algorithm in \cite{Goldsmith} under the user locations presented in Fig. \ref{FIG_Snapshot_Beam}.  ``$\circ$'' represents a user. Users and beams in the same color belong to the same subnetwork. Inactive beams are drawn in dashed lines. $N=32$, $K=10$.}
	\label{FIG_Snapshot_Beam_Gold}
\end{figure*}

To apply the AP-centric clustering algorithm in \cite{Goldsmith} in a single-cell massive MIMO system, distances between beams need to be defined. As inter-beam interference decreases with the angular separation between two beams, and the first beam and the last beam are adjacent, the distance between any beam $b_{i}$ and beam $b_{j}$ can be set as $\min\left\{\cos\theta_{i}-\cos\theta_{j}, \cos\theta_{j}-\cos\theta_{i}+2\right\}$ with $\theta_{i}\leq \theta_{j}$, where $\theta_{i}$ is the main direction of beam $b_{i}$. Fig. \ref{FIG_Snapshot_Beam_Gold} presents the decomposition results with the AP-centric clustering algorithm in \cite{Goldsmith} by varying the number of decomposed subnetworks $M$. It can be seen that different subnetworks contain a similar amount of beams, which limits the flexibility to group beams and users.  
 
\begin{figure*}[t]
 	\centering
 	{	\subfloat[]{\includegraphics[width=0.45\textwidth]{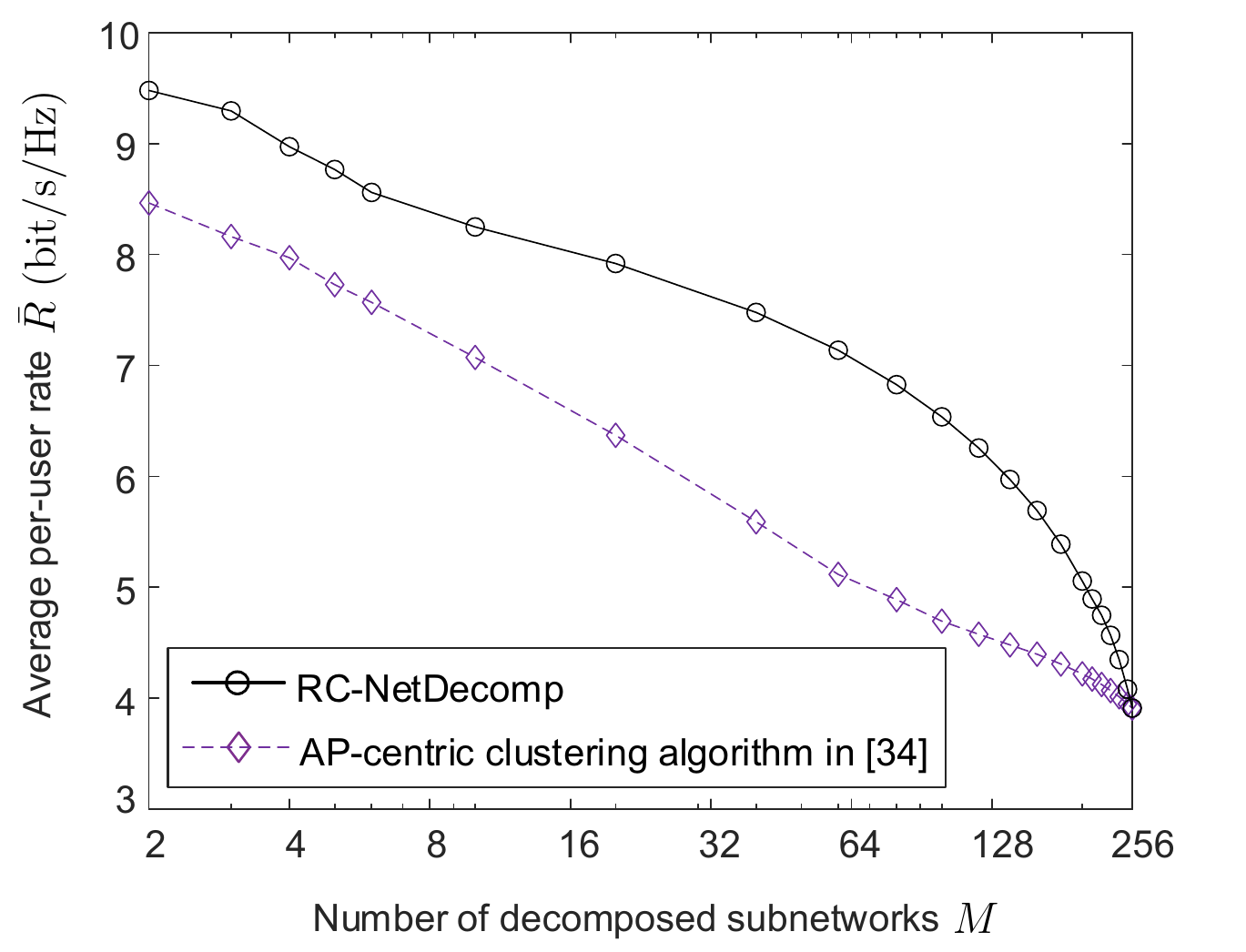}
 			\label{FIG_AveR_Beam_all}}
 		\subfloat[]{\includegraphics[width=0.45\textwidth]{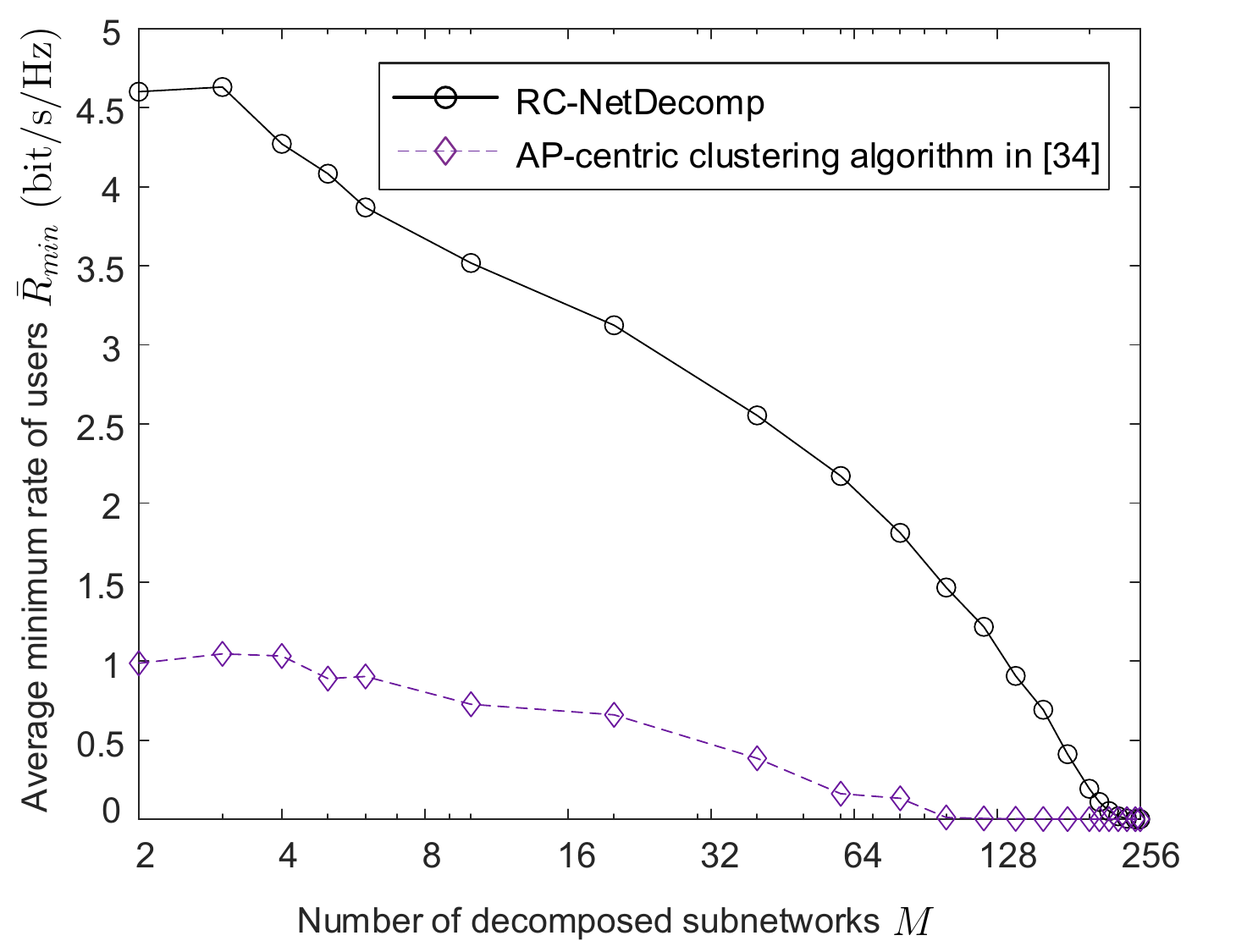}
 			\label{FIG_AveRmin_Beam_all}} \\
 		\subfloat[]{\includegraphics[width=0.45\textwidth]{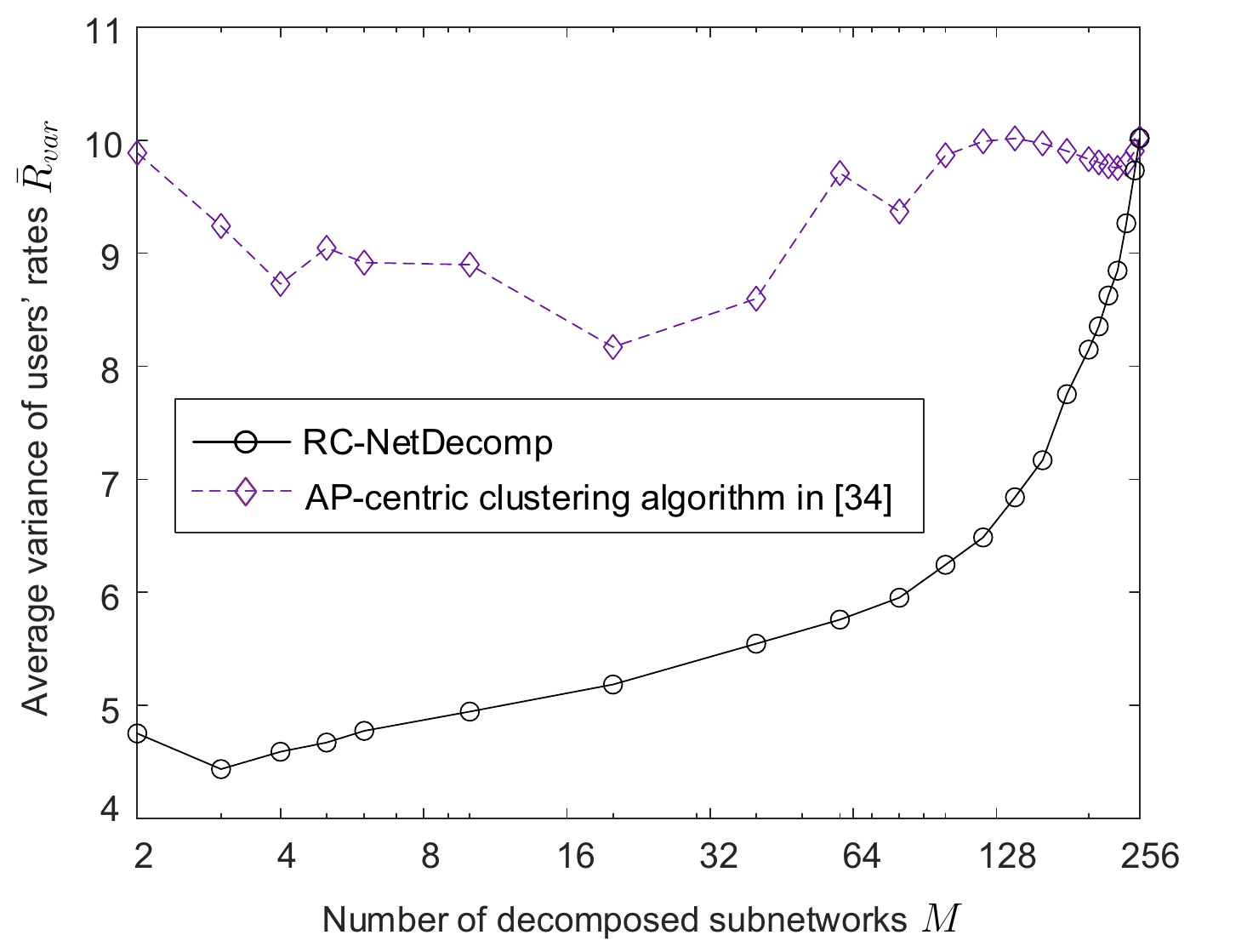}
 			\label{FIG_AveRvar_Beam_all}}	
 		\subfloat[]{\includegraphics[width=0.45\textwidth]{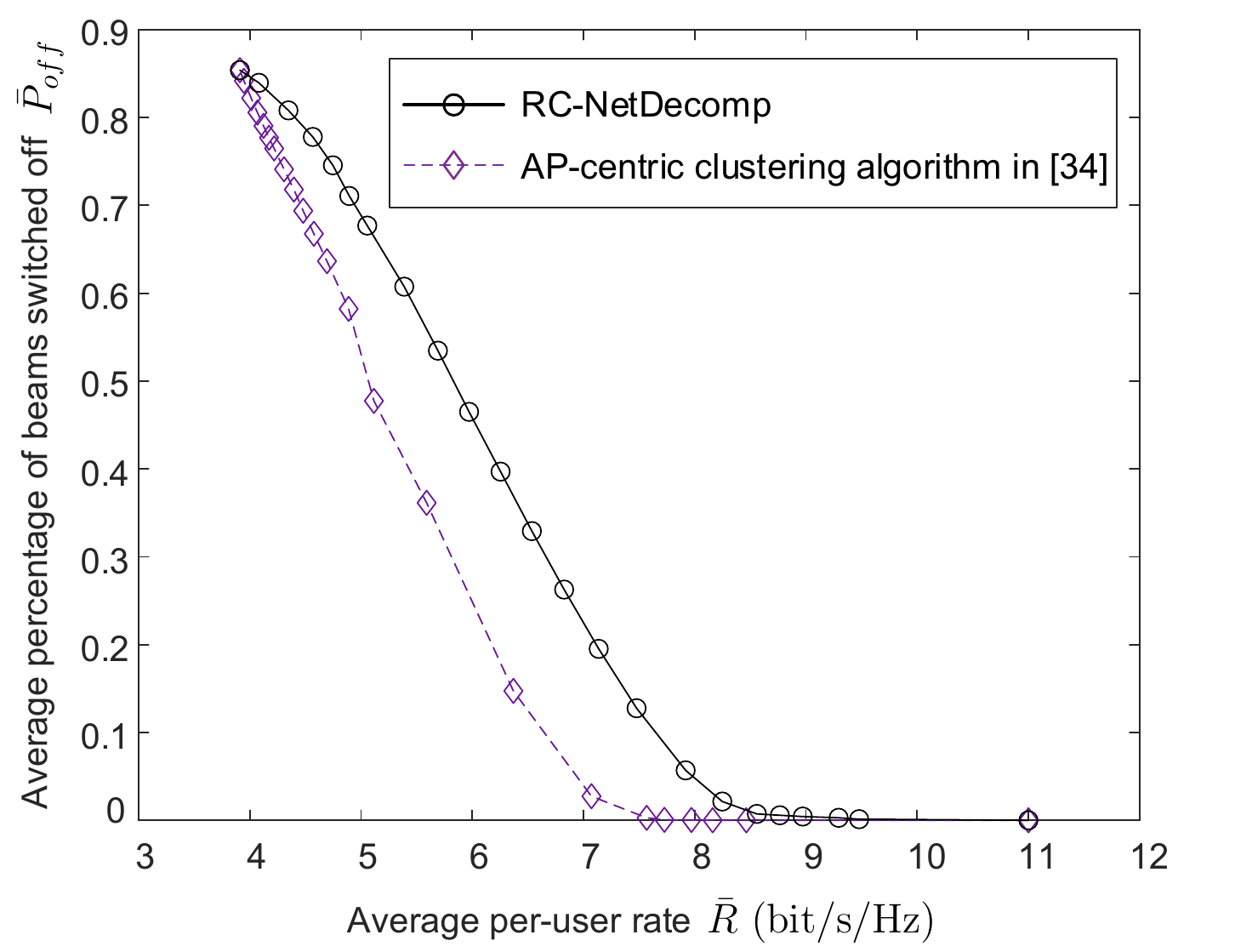}
 			\label{FIG_AveIsoAp_Beam_all}}
 	}
 	\caption{(a) Average per-user rate $\bar{R}$, (b) average minimum rate of users $\bar{R}_{min}$, (c) average variance of users' rates $\bar{R}_{var}$ and (d) average percentage of beams switched off $\bar{P}_{off}$ with the proposed RC-NetDecomp algorithm and the AP-centric clustering algorithm  in \cite{Goldsmith}. $N =256$, $K=50$, $P_{t}/\sigma^2=0$dB, $\alpha=2.7$.}
 	\label{FIG_Ave_Beam}
\end{figure*}

Figs. \ref{FIG_AveR_Beam_all}--\ref{FIG_AveRvar_Beam_all} present the average per-user rate $\bar{R}$, the average minimum rate of users $\bar{R}_{min}$ and the average variance of users' rates  $\bar{R}_{var}$ versus the number of decomposed subnetworks $M$, and Fig. \ref{FIG_AveIsoAp_Beam_all} presents the average percentage of beams switched off  $\bar{P}_{off}$ versus the average per-user rate  $\bar{R}$ with both the proposed RC-NetDecomp algorithm and the AP-centric clustering algorithm in \cite{Goldsmith}. It can be seen that the proposed RC-NetDecomp algorithm outperforms the AP-centric clustering algorithm in terms of average per-user rate, user fairness by achieving a higher average minimum rate of users  $\bar{R}_{min}$ and a lower variance of users' rates  $\bar{R}_{var}$, and energy efficiency by switching off more beams while achieving the same average per-user rate. Note that in Fig. \ref{FIG_AveRvar_Beam_all}, fluctuations are observed in the average variance of users’ rates with the AP-centric algorithm in \cite{Goldsmith}. This is because the algorithm in \cite{Goldsmith} was originally proposed for cell-free massive MIMO systems, where the distance measure for clustering is Euclidean distance. Here, the Euclidean distances between APs are replaced with the angular separations between beams to implement it in single-cell massive MIMO systems, whereas the beam gain from a beam to a user fluctuates as their angular separation increases.

\vspace{-0.1mm}
\section{Conclusion}
This paper focused on the rate-constrained clustered cell-free networking problem in a wireless network with multiple APs, where multiple beams are available at each AP. With the aim of maximizing the number of decomposed subnetworks under a per-user rate constraint, the clustered cell-free networking problem was formulated as a bipartite graph partitioning problem and a RC-NetDecomp algorithm was proposed. Since the edge weight between a user and a beam was defined as the corresponding normalized channel gain, the proposed RC-NetDecomp algorithm can switch off the beams that are badly aligned with users or associated with APs far from users to save energy, and produce subnetworks with balanced sizes at the same time. The performance of the proposed RC-NetDecomp algorithm was evaluated and compared against the user-centric and AP-centric benchmarks in two special cases of cell-free massive MIMO and single-cell massive MIMO. Simulation results showed that in both cases, the proposed RC-NetDecomp algorithm provides a fine-grained tuning of subnetwork size and produces more balanced subnetworks with smaller maximum subnetwork size than the user-centric baseline, which can efficiently reduce the joint processing complexity and signaling overhead in practice. Compared to the AP-centric baseline, higher average per-user rate, better user fairness and higher energy efficiency can be achieved by adopting the proposed RC-NetDecomp algorithm. 

Note that to implement the RC-NetDecomp algorithm, channel gains between beams and users in the network need to be collected. To reduce the channel measurement overhead, it is of paramount importance to further develop distributed clustered cell-free networking schemes in the future, where each AP only needs to measure the channels of the users within its vicinity and exchange information with a small number of nearby APs.

\numberwithin{equation}{section}

\appendices

\section{Derivation of (\ref{Prob-1-new})--(\ref{const-inter-Cm-sub1-new})}
Recall that a new graph $\tilde{G}=(\tilde{\mathcal{V}}, \tilde{E})$ is constructed by merging each beam vertex $b_{i}$ with the users vertices with edge weight 1 between them  on graph $G$ to form the meganode $\tilde{v}_{i}$ given in (\ref{meganode}). Suppose that graph $\tilde{G}$ is partitioned into $M$ subgraphs and the set of meganodes in the $m$th subgraph is denoted as $\tilde{\mathcal{C}}_{m}$ with $\bigcup_{m=1}^{M}\tilde{\mathcal{C}}_{m}=\tilde{\mathcal{V}}$ and $\tilde{\mathcal{C}}_{m}\cap\tilde{\mathcal{C}}_{m^{'}}=\emptyset, \forall m^{'}\neq m$. The corresponding set of user and beam vertices in the $m$th subgraph on graph $G$, $C_{m}$,  is then given by
\begin{align}\label{Cm'}
C_{m}=\bigcup_{\tilde{v}_{i}\in \tilde{C}_{m}}\tilde{v}_{i}=\left\{b_{i}: \tilde{v}_{i}\in \tilde{\mathcal{C}}_{m}\right\}\cup \left\{u_{k}: u_{k}\in \tilde{v}_{i}, \tilde{v}_{i}\in \tilde{\mathcal{C}}_{m} \right\}.
\end{align}
By combining (\ref{cut}), (\ref{meganode}) and (\ref{Cm'}), the cut function $\text{cut}(C_{m})$ can be rewritten as
\begin{align}\label{cut-new}
\text{cut}(C_{m})&{=}\hspace{-2mm}\sum_{u_{k}\in C_{m}}\sum_{b_{n}\in B, b_{n}\notin C_{m}} w_{k,n}+\sum_{b_{n}\in C_{m}}\sum_{u_{k}\in U, u_{k}\notin C_{m}} w_{k,n} \nonumber \\
&{=}\hspace{-2mm}\sum_{\tilde{v}_{i}\in \tilde{\mathcal{C}}_{m}}\sum_{u_{k}\in \tilde{v}_{i}}\sum_{\tilde{v}_{j}\in \tilde{\mathcal{V}}, \tilde{v}_{j}\notin \tilde{\mathcal{C}}_{m}} \hspace{-4mm} w_{k,j}{+}\hspace{-2mm}\sum_{\tilde{v}_{i}\in \tilde{\mathcal{C}}_{m}}\sum_{\tilde{v}_{j}\in \tilde{\mathcal{V}}, \tilde{v}_{j}\notin \tilde{\mathcal{C}}_{m}}\sum_{u_{k}\in \tilde{v}_{j}}\hspace{-1mm} w_{k,i}
{=}\hspace{-2mm}\sum_{\tilde{v}_{i}\in \tilde{\mathcal{C}}_{m}}\sum_{\tilde{v}_{j}\in \tilde{\mathcal{V}}, \tilde{v}_{j}\notin \tilde{\mathcal{C}}_{m}}\hspace{-2mm}\left(\sum_{u_{k}\in \tilde{v}_{i}} \hspace{-2mm}w_{k,j}{+}\sum_{u_{k}\in \tilde{v}_{j}} \hspace{-2mm}w_{k,i} \right). 
\end{align}
By substituting (\ref{edge-new}) into (\ref{cut-new}), we have
\begin{align}\label{cut-cut-new}
	\text{cut}(C_{m})=\sum_{\tilde{v}_{i}\in \tilde{\mathcal{C}}_{m}}\sum_{\tilde{v}_{j}\in \tilde{\mathcal{V}}, \tilde{v}_{j}\notin \tilde{\mathcal{C}}_{m}} \tilde{w}_{i,j}=\text{cut}(\tilde{\mathcal{C}}_{m}).
\end{align}
As a meganode on the new graph $\tilde{G}$ always contains one beam, constraint (\ref{const-iso-user-sub1-cut}) is always satisfied. According to (\ref{cut-cut-new}), the equivalent graph partitioning problem on graph $\tilde{G}=(\tilde{\mathcal{V}}, \tilde{E})$ given in (\ref{Prob-1-new})--(\ref{const-inter-Cm-sub1-new}) can be therefore obtained from (\ref{Prob-1-cut})--(\ref{const-iso-user-sub1-cut}). 

\vspace{-2mm}
\section{Proof of Theorem 1}
\begin{IEEEproof}
	The mincut function $\sum_{m=1}^{M+1}\text{cut}(C_{m|M+1}^{*})$ for a given number of subgraphs $M+1$ can be written as
	\begin{align}\label{ApB-M}
	\sum_{m=1}^{M+1}\text{cut}(C_{m|M+1}^{*})=\sum_{m=1}^{M-1}\text{cut}(C_{m|M+1}^{*})+\text{cut}(C_{M|M+1}^{*})+\text{cut}(C_{M+1|M+1}^{*}).
	\end{align} 
	Recall that the cut function of a set $X$, $\text{cut}(X)$, is given in (\ref{def-cut}) as $\text{cut}(X)=\sum_{v_{i}\in X}\sum_{v_{j}\in \overline{X}} a_{i,j}$.
	%\begin{align}\label{ApB-cut}
	%\text{cut}(X)=\sum_{v_{i}\in X}\sum_{v_{j}\in \overline{X}} a_{i,j}.
	%\end{align}
	For any two sets $X, Y \subseteq V$ and $X\cap Y=\emptyset$, the cut function, $\text{cut}(X \cup Y)$, can be obtained as
	\begin{align}\label{ApB-cutxy}
	\text{cut}(X \cup Y)=\sum_{v_{i}\in X\cup Y}\sum_{v_{j}\in \overline{X\cup Y}} a_{i,j}=\sum_{v_{i}\in X}\sum_{v_{j}\in \overline{X\cup Y}} a_{i,j}+\sum_{v_{i}\in Y}\sum_{v_{j}\in \overline{X\cup Y}} a_{i,j}.
	\end{align}
	For any set $X$, it is obvious that $\overline{X\cup Y} \subseteq \overline{X}$.
	%\begin{align}\label{ApB-x}
	%\overline{X\cup Y} \subseteq \overline{X}.
	%\end{align}
	%By combining (\ref{ApB-cutxy}) and (\ref{ApB-x}), we have
	We then have
	\begin{align}\label{ApB-cutxy-final}
	\text{cut}(X \cup Y)\leq \sum_{v_{i}\in X}\sum_{v_{j}\in \overline{X}} a_{i,j}+\sum_{v_{i}\in Y}\sum_{v_{j}\in \overline{Y}} a_{i,j}=\text{cut}(X)+\text{cut}(Y).
	\end{align}
	The mincut function $\sum_{m=1}^{M+1}\text{cut}(C_{m|M+1}^{*})$ given in (\ref{ApB-M}) is then lower-bounded by
	\begin{align}\label{ApB-M-lb}
	\sum_{m=1}^{M+1}\text{cut}(C_{m|M+1}^{*})\geq \sum_{m=1}^{M-1}\text{cut}(C_{m|M+1}^{*})+\text{cut}(C_{M|M+1}^{*}\cup C_{M+1|M+1}^{*}) \geq \sum_{m=1}^{M}\text{cut}(C_{m|M}^{*}),
	\end{align}
	as $\mathcal{M}_{|M}^{*}=\left\{C_{1|M}^{*}, C_{2|M}^{*}, \cdots C_{M|M}^{*}\right\}$ is the optimal $M$-way partition that minimizes the cut function $\sum_{m=1}^{M}\text{cut}(C_{m})$.
\end{IEEEproof}

%\linespread{1.41}

\end{document}